\documentclass[11pt, a4paper]{article}
\pdfoutput=1
\usepackage{jheparxiv}
\usepackage[latin1]{inputenc}
\usepackage{amsmath}
\usepackage{amsfonts}
\usepackage{amssymb}
\usepackage{latexsym}
\usepackage{mathrsfs}
\usepackage{graphicx}
\usepackage{color}
\usepackage{slashed}
\usepackage{twistor}

\def\bd {\boldsymbol}

\subheader{\hfill \texttt{DAMTP-2013-74}}

\title{Ambitwistor strings and the scattering equations at one loop}

\author{Tim Adamo, Eduardo Casali and David Skinner}

\affiliation{Department of Applied Mathematics \& Theoretical Physics \\
        University of Cambridge \\
        Wilberforce Road \\
        Cambridge CB3 0WA, United Kingdom}

\emailAdd{[t.adamo, e.casali, d.b.skinner]@damtp.cam.ac.uk}

\abstract{Ambitwistor strings are chiral, infinite tension analogues of conventional string theory whose target space is the space of complex null geodesics and whose spectrum consists exclusively of massless states. At genus zero, these strings underpin the Cachazo--He--Yuan formul\ae\ for tree level scattering of gravitons, gluons and scalars. In this paper we extend these formul\ae\ in a number of directions. Firstly, we consider Ramond sector vertex operators and construct simple amplitudes involving space--time fermions. These agree with tree amplitudes in ten dimensional supergravity and super Yang--Mills.  We then show that, after the usual GSO projections, the ambitwistor string partition function is modular invariant. We consider the scattering equations at genus one, and calculate one loop scattering amplitudes for NS--NS external states in the Type II ambitwistor string. We conjecture that these give new representations of (the integrand of) one loop supergravity amplitudes and we show that they have the expected behaviour under factorization of the worldsheet in both non--separating and separating degenerations.}

\begin{document}

\maketitle

\section{Introduction}
\label{sec:intro}

Recently, a new class of worldsheet theories called \emph{ambitwistor strings} were discovered \cite{Mason:2013sva}.  These are chiral, infinite tension analogues of ordinary string theory, whose basic worldsheet action takes the first--order form
\be\label{wsai}
\frac{1}{2\pi}\int_{\Sigma} P_{\mu}\dbar X^{\mu}-\frac{e}{2} P^{2}
\ee
in a natural generalization of the usual worldline action for a massive particle. The fields $P$ and $X$ represent holomorphic coordinates on the cotangent bundle of complexified space--time. However, to be able to integrate the worldsheet Lagrangian over $\Sigma$, $P_\mu$ must also be a worldsheet (1,0)-form. Consequently, the field $e$ is a Beltrami differential that imposes the constraint that the quadratic differential $P^2=P_\mu P^\mu$ vanishes. The action~\eqref{wsai} has a gauge redundancy $\delta X^\mu = \alpha P^\mu$ that is conjugate to this constraint, which implies that field configurations that differ by translation along a null geodesic in space--time are to be considered equivalent. Together, the constraint and gauge redundancy mean that the target space is properly understood to be the space of complex null geodesics, known as (projective) ambitwistor space $\PA$ \cite{LeBrun:1983, Witten:1985nt}.

In~\cite{Mason:2013sva} it was shown that these ambitwistor strings have no massive states in their spectrum, essentially on account of the triviality of the $XX$ OPE. Like the usual string, the bosonic model~\eqref{wsai} admits both supersymmetric and heterotic generalizations. In particular, the NS--NS sector spectrum of a Type II ambitwistor string was shown to agree with the Neveu--Schwarz sector of ten dimensional supergravity, with no $\alpha'$ corrections, while the heterotic model also involved a coupling to ten dimensional Yang--Mills theory. A (gauge--fixed) pure spinor version of the ambitwistor string was subsequently constructed in~\cite{Berkovits:2013xba}.

\medskip

The discovery of the ambitwistor string was motivated by trying to understand the origin of the representations of the tree level S--matrices of gravity and Yang--Mills obtained in~\cite{Cachazo:2013hca,Cachazo:2013iea}. See~\cite{Dolan:2013isa} for a proof of the Yang--Mills formula via BCFW methods. The key feature of these representations is that, while (as in string theory) the amplitudes are written in terms of an integral over the moduli space of an $n$-pointed Riemann sphere, this integral is completely localized to solutions of the genus zero {\it scattering equations}
\be\label{int3}
	\sum_{j\neq i} \frac{k_i\cdot k_j}{z_i-z_j} = 0 \qquad\hbox{for} \qquad i\in\{1,\ldots,n\!-\!4\}\,.
\ee
These equations were first discovered by Gross \& Mende~\cite{Gross:1987kza,Gross:1987ar} who showed that they dominate the behaviour of usual string theory in the limit of high energy scattering at fixed angle. They also appear in the context of twistor strings, as shown in~\cite{Witten:2004cp}.

The scattering equations have an extremely natural interpretation in the ambitwistor string context. In the presence of vertex operator insertions, the field $P_\mu$ is not globally holomorphic on $\Sigma$, but rather has poles whose residues are determined by the external momenta. Likewise, $P^2$ becomes a {\it meromorphic} quadratic differential with poles only at the vertex operator insertion points. At genus zero, the space of such meromorphic quadratic differentials is $n\!-\!3$ dimensional. Now, noting that $P^2$ vanishes so that $P_\mu$ itself is a (complex) null vector, was a crucial step in deducing that the target space of~\eqref{wsai} is indeed ambitwistor space and that the gauging of $\delta X^\mu = \alpha P^\mu$ is well--defined. The scattering equations simply enforce that the residues of $P^2$ vanish at any $n\!-\!3$ of the insertion points, and hence that $P^2$ indeed vanishes globally on $\Sigma$.

\medskip

More than just providing the underlying geometric explanation of the formul\ae\ of~\cite{Cachazo:2013iea}, the ambitwistor string allows us to extend these formul\ae\ in a variety of ways. In this paper, we consider two main extensions: to scattering amplitudes in supergravity and super Yang--Mills (including fermionic states) and to loop amplitudes. After reviewing the construction of the ambitwistor string in section~\ref{sec:review}, we construct Ramond sector vertex operators representing space--time gravitinos and, in the heterotic model, gauginos. The simplest tree level amplitudes involving these fermionic states are computed in section~\ref{sec:fermions}, and are shown to agree with known expressions for gravitino and gaugino scattering amplitudes in supergravity and super Yang--Mills. 

In section~\ref{sec:1loop} we consider the ambitwistor string at genus one. We show that the partition function is modular invariant --- a non--trivial constraint on a purely chiral theory --- and construct the correct generalization of the scattering equations~\eqref{int3} to elliptic curves. We find that as well as imposing conditions on the residues of $P^2$ at $n-1$ of the vertex operators, we need a further condition on $P^2$ itself. This just reflects the fact that the moduli space for genus one curves includes a specification of the complex structure $\tau$ of the torus as well as a choice of the $n$ marked points, up to an overall translation around the torus. In section~\ref{sec:1loopamps} we compute $n$-point correlation functions of gravitational vertex operators on our genus 1 curve. As at genus zero~\cite{Cachazo:2013hca,Mason:2013sva}, these take the form of Pfaffians whose dependence on the worldsheet coordinates is fixed by the $g=1$ scattering equations. 

We conjecture that these formul\ae\ are new representations of $n$-point gravitational one loop scattering amplitudes in ten dimensional Type II supergravity (IIA or IIB according to the choice of GSO projection). In particular, in section~\ref{sec:factorize}, we check that our amplitudes have the expected behaviour in both non--separating and separating degenerations of the worldsheet, corresponding to the single cut of the loop amplitude and to factorization of a tree sub--amplitude from off the loop, respectively. These checks demonstrate that the worldsheet theory factorizes unitarily at higher genus, and also provide evidence in favor of interpreting the one-loop expressions as gravitational amplitudes. Crucially, the result of the single cut is a {\it rational} function of the kinematic data (for the same reason that the tree-level amplitudes are), to be integrated over the on-shell phase space of the intermediate particle.  This is as expected for field theoretic amplitudes, and stands in contrast to what happens in a generic string theory where an infinite tower of string oscillators propagate around the loop.

We conclude in section~\ref{sec:conclusions} with a brief look at some open questions.


\section{The ambitwistor string}
\label{sec:review}

We begin with a brief review of ambitwistor string theory, focussing on the type II model associated with gravity.  After first reviewing the worldsheet theory and its BRST symmetries, we discuss the structure of the Neveu-Schwarz and Ramond sector vertex operators which describe gravitons, dilatons, $B$-fields, and space-time fermions.  Further details may be found in~\cite{Mason:2013sva}.

\subsection{Type II worldsheet theory}

The worldsheet action for the type II ambitwistor string introduced in~\cite{Mason:2013sva} is 
\be
	S = \frac{1}{2\pi}\int_\Sigma P_\mu\delbar X^\mu - \frac{1}{2} e P_\mu P^\mu + \Psi_\mu\delbar\Psi^\mu  - \chi P_\mu\Psi^\mu + \tilde\Psi_\mu\delbar\tilde\Psi^\mu - \tilde\chi P_\mu\tilde\Psi^\mu
\label{S}
\ee
where $P_\mu\in\Omega^{1,0}(\Sigma)$ and so may be written as $P_\mu = P_{\mu z}\rd z$ in terms of some local holomorphic coordinate $z\in\Sigma$.  Both $\Psi_{\mu}$ and $\tilde{\Psi}_{\mu}$ are worldsheet fermions of the \emph{same} chirality, which are elements of $\Pi\Omega^{0}(\Sigma, K^{1/2}_{\Sigma})$, where $\Pi$ is the parity reversing functor.  Consequently, for \eqref{S} to be well-defined we must have $e\in\Omega^{0,1}(\Sigma, T_{\Sigma})$ and $\chi,\tilde\chi\in\Omega^{0,1}(\Sigma, T^{1/2}_{\Sigma})$. 

This action has gauge redundancies that act as
\be
\begin{aligned}
	\delta X^\mu & = \alpha P^\mu + \epsilon\Psi^\mu + \tilde\epsilon\tilde\Psi^\mu \qquad\qquad &\delta P_\mu &= 0\\
	\delta\Psi^\mu &= \epsilon P^\mu &\delta\tilde\Psi^\mu &=\tilde\epsilon P^\mu\\
\end{aligned}
\label{gauge1}
\ee
on the matter fields and
\be
	\delta e = \delbar\alpha \qquad\qquad \delta \chi = \delbar\epsilon\qquad\qquad \delta\tilde\chi = \delbar\tilde\epsilon
\label{gauge2}
\ee
on the gauge fields, where $\alpha\in\Omega^0(\Sigma,T_\Sigma)$ is bosonic while $\epsilon,\tilde\epsilon\in\Omega^0(\Sigma,T_\Sigma^{1/2})$ are fermionic. In particular, as emphasized in~\cite{Mason:2013sva}, the bosonic gauge field $e$ imposes the constraint that $P_\mu$ is null with respect to the target space metric and the associated transformation $\delta X^\mu = \alpha P^\mu$ instructs us to count as equivalent field configurations that differ only by translation of $X$ along a null direction.  Hence, the target space is properly understood as the space of (complex) null geodesics, or ambitwistor space.

As usual, we can gauge--fix the chiral worldsheet gravity and gravitinos of this theory by introducing a $bc$-ghost system and two copies of the superconformal ghost system which we denote as $\beta\gamma$ and $\tilde{\beta}\tilde{\gamma}$.  In particular, these fields take values in the usual spaces
\be
\begin{aligned}
	b\in\Pi\Omega^{0}(\Sigma,K^{2}_{\Sigma}) \qquad \beta,\tilde{\beta}\in\Omega^{0}(\Sigma,K^{3/2}_{\Sigma}) \\
	c\in\Pi\Omega^{0}(\Sigma, T_{\Sigma}) \qquad \gamma,\tilde{\gamma}\in\Omega^{0}(\Sigma, T^{1/2}_{\Sigma})\,,
\end{aligned}
\ee
except that both sets of ghost systems are chiral (left--moving). The only difference from the gauge--fixing of ordinary string theory is the need to fix the gauge redundancy associated to $\alpha$ in \eqref{gauge1}--\eqref{gauge2}. To do this, we follow the usual BRST procedure and add to the action a gauge--fixing term
\be\label{gaugefix}
\left\{ Q, \int_\Sigma \tilde b \; F(e) \right\},
\ee
where $\tilde{b}\in\Pi\Omega^{0}(\Sigma,K^2_{\Sigma})$ and $F(e)$ is a gauge--fixing functional.  Naturally, we would like to choose $F$ to set $e=0$; the obstruction to doing this is given by the moduli of the problem.  In particular, the BRST transformations of the gauge fields only allow us to vary $e$ within a fixed Dolbeault cohomology class.  If $\Sigma$ is a genus $g$ Riemann surface with $n$ marked points $\{z_i\}$ at which the gauge transformation is required to vanish, then for $r=1,\ldots,3g-3+n$ let $\{\mu_r\}$ be a basis of $H^{0,1}(\Sigma, T_{\Sigma}(-z_1-\cdots-z_n))$.  We can then choose our gauge--fixing functional to be:
\be\label{gff}
F(e)=e-\sum_{r=1}^{3g-3+n} s_{r}\;\mu_{r},
\ee
where $s_r\in\C$ are coefficients of the basis.

Now, the action of the BRST operator $Q$ on the various fields in the gauge--fixing term is
\begin{equation*}
 \delta\tilde{b}=m, \qquad \delta e=\dbar\tilde{c}, \qquad \delta s_{r}=q_{r}, \qquad \delta m =0,  \qquad \delta q_{r}=0 ,
\end{equation*}
so after integrating out the Lagrange multiplier $m$, the relevant part of the action \eqref{S} becomes
\be
\frac{1}{2\pi}\int_{\Sigma} \tilde{b}\;\dbar\tilde{c}-\sum_r s_{r}\int_\Sigma \mu_{r}\,P^{2}-\sum_{r=1}^{3g-3+n}q_{r}\int_\Sigma \tilde{b}\,\mu_{r}\ .
\ee
Integrating out the bosonic and fermionic parameters $s_r$ and $q_r$ leaves us with an insertion of
\be\label{tildemoduli}
\prod_{r=1}^{3g-3+n}\bar{\delta}\left(\int_{\Sigma}\mu_rP^{2}\right)\;\int_{\Sigma}\tilde{b}\,\mu_{r}
\ee
inside the path integral.

As in the holomorphic worldsheet reparametrizations of ordinary string theory, at genus zero we can choose a basis of our $n\!-\!3$ Beltrami differentials so that $\int\tilde b\,\mu_r$ simply extracts the residue of $\tilde b$ at the location of the $r^{\rm th}$ vertex operator. This then strips off the $\tilde c$ ghost associated with a (fixed) vertex operator insertion. Similarly, the integral $\int \mu_r P^2$ in~\eqref{tildemoduli} extracts the residue of the quadratic differential $P^2$ at the location of the vertex operator, leaving us with a $\delta$-function that forces this residue to vanish. At genus zero, a quadratic differential must have at least four poles (counted with multiplicity). Below, we shall see that $P^2$ has at most simple poles, so enforcing vanishing of all but three of its residues ensures that in fact $P^2=0$ globally over the genus zero Riemann surface. This is exactly the content of the scattering equations~\cite{Gross:1987kza}. They emerge here as a natural consequence of the gauge redundancy enforcing that the target space is ambitwistor space in the presence of vertex operator insertions.

Finally, we are left with the gauge--fixed action~\cite{Mason:2013sva}
\be\label{gfa}
S=\frac{1}{2\pi}\int_{\Sigma}P_\mu\delbar X^\mu + \Psi_\mu\delbar\Psi^\mu + \tilde\Psi_\mu\delbar\tilde\Psi^\mu +b\;\dbar c+\tilde{b}\;\dbar\tilde{c}+\beta\dbar\gamma +\tilde{\beta}\dbar \tilde{\gamma}.
\ee 
describing type II ambitwistor strings. Requiring that this theory have vanishing central charge restricts the space--time dimension to be $d=10$, so the critical dimension of the type II ambitwistor string is the same as the type II superstring. Although this theory will be our focus for most of this paper, it should be mentioned that there is a heterotic version of the ambitwistor string.  This is nearly identical to the type II model given here, except that the $\tilde{\Psi}$ system is exchanged for a worldsheet current algebra for some gauge group.  At genus zero and leading trace in the current algebra this describes vector bosons of the chosen gauge group \cite{Mason:2013sva}, although interactions of the gravitational degrees of freedom in the heterotic model are not understood.

\subsection{Neveu--Schwarz sector vertex operators}

The basic NS--NS vertex operator in the type II model is
\be\label{fixedVO}
	c\tilde c \; U(z) = c\tilde c \, \delta(\gamma)\,\delta(\tilde\gamma)\,\epsilon \cdot\Psi\ \tilde\epsilon\cdot\tilde\Psi \; \e^{\im k\cdot X(z)},
\ee
where $\epsilon$, $\tilde{\epsilon}$ are the polarizations and $k$ is a momentum vector. The antisymmetric, symmetric trace--free and trace parts of $\epsilon_\mu\tilde\epsilon_\nu$ represent a $B$-field, graviton and dilaton, respectively. The form of this vertex operator is thus practically identical to that of fixed NS vertex operators in type II string theory; the only difference is that all the fields in the ambitwistor string are chiral and have only holomorphic conformal weight. Note that the total conformal weight vanishes, as it must. Another important difference to the usual string is that~\eqref{fixedVO} is the only vertex operator present in the NS--NS sector after imposing the usual GSO projection $(-1)^{\rm F} = (-1)^{\tilde{\rm F}} = +1$ on both the $\Psi$ and $\tilde\Psi$ systems. This is because the $XX$ OPE is trivial, so in particular $\e^{\im k\cdot X}$ has vanishing (anomalous) conformal weight.

As usual, the insertion of such vertex operators is really an instruction to quotient the path integral only by those gauge transformations that vanish at the insertion points. Following the standard descent procedure for the worldsheet supersymmetries (see {\it e.g.}~\cite{Polchinski:1998rr, Witten:2012bh}) transforms~\eqref{fixedVO} into a vertex operator
\be\label{fixedV}
	c \tilde{c}\;V=c\tilde{c}\;\left(\epsilon\cdot P+k\cdot\Psi\;\epsilon\cdot\Psi\right)\left(\tilde{\epsilon}\cdot P+k\cdot\Psi\;\tilde{\epsilon}\cdot\Psi\right)\;\e^{\im k\cdot X},
\ee
which is inserted at a fixed location on $\Sigma$, but no longer fixes a zero mode of the superconformal ghosts $\gamma$, $\tilde\gamma$.  This composite operator is well--defined provided $k^2=0$ and $\epsilon\cdot k = \tilde\epsilon\cdot k =0$. It is BRST trivial if $\epsilon$ or $\tilde\epsilon$ are proportional to $k$, giving the usual linearized diffeomorphism invariance in space--time.

Finally, we may obtain a vertex operator that is integrated over $\Sigma$ by pairing~\eqref{fixedV} with moduli insertions from the gauge--fixing as
\begin{equation*}
 \left(\int_{\Sigma}b\wedge\mu_r\right)\left(\int_{\Sigma}\tilde{b}\wedge\mu_{r}\right)\ \bar{\delta}\left(\int_{\Sigma}\mu_r\,P^{2}\right)\;\; c\tilde{c}\;V(z).
\end{equation*}
As usual, the factors of $b$ and $\tilde b$ in the measure have the effect of removing the ghost factor $c\tilde c$ from the vertex operator. As above, the remaining $\bar\delta\left(\int P^2 \mu_r\right)$ factor imposes that the residue of the quadratic differential $P^2$ should vanish at the insertion point. To compute this residue, notice that together with the kinetic term $\frac{1}{2\pi}\int_\Sigma P_\mu\delbar X^\mu$ in the action, these vertex operators provide the only $X$ dependence in the path integral. Integrating out the zero modes of $X$ leads to an overall momentum conserving $\delta$-function, while integrating out the non--zero modes leads to the constraint
\be
\label{intoutX}
	\delbar P_\mu = 2\pi\im\,\rd z\wedge\rd \bar z \,\sum_{i=1}^n k_i \,\delta^2(z-z_i) 
\ee
on the 1-form $P_\mu$. Thus $P_\mu$ is holomorphic everywhere except at the insertion points, where is has residue given by the external momentum $k_i$. Since all these external momenta obey $k_i^2=0$, $P^2$ also has only simple poles and\footnote{Here we understand the residue at a point $z_i$ of a quadratic differential  on $\Sigma$ to be a 1-form at $z_i$. This follows from the residue exact sequence
$$
	0 \to K^2 \hookrightarrow K^2(z_i) \buildrel{\rm Res}\over\longrightarrow \left.K\right|_{z_i}\to 0
$$
on $\Sigma$, where the second map is inclusion and the third is the residue map at $z_i$. See {\it e.g.}~\cite{Witten:2004cp} for further explanation in a similar context.} ${\rm Res}_i\, P^2(z)= k_i\cdot P(z_i)$. Thus we are left with an integrated vertex operator
\be\label{intVO}
	\int_{\Sigma}\bar{\delta}\!\left(k\cdot P(z)\right)V(z)\,.
\ee
The integral makes sense because $V$ is a quadratic differential on $\Sigma$, while $\bar{\delta}(k\cdot P)$ takes values in $\Omega^{0,1}(\Sigma,T_{\Sigma})$.

Using the ambitwistor version of the Penrose transform~\cite{LeBrun:1983,Baston:1987av} it can be shown that the vertex operators \eqref{fixedVO}--\eqref{intVO} represent the NS--NS sector of supergravity in ten dimensions~\cite{Mason:2013sva}.  They may be contrasted against the momentum eigenstates used in twistor space for four dimensional flat space--time (see {\it e.g}. \cite{Witten:2004cp, Adamo:2011pv})
\begin{equation*}
 	\bar{\delta}\left(\la \lambda\, \lambda(z)\ra\right)\;\e^{\im [\mu(z)\tilde{\lambda}]}\,,
\end{equation*}
where $p^{\alpha\dot{\alpha}}=\lambda^{\alpha}\tilde{\lambda}^{\dot{\alpha}}$ is an on--shell four--momentum and $Z^{A}(z)=(\lambda_{\alpha},\mu^{\dot{\alpha}})(z)$ are homogeneous coordinates on twistor space.  In twistor space, the mass--shell condition is encoded automatically, but vertex operators of different homogeneity are needed to describe different helicity states.  The ambitwistor wavefunction 
\begin{equation*}
	\bar{\delta}\left(k\cdot P(z)\right)\;\e^{\im k\cdot X(z)}
\end{equation*}
that appears in~\eqref{intVO} can be seen as an analogue of the twistor wavefunction. The fact that $k^2=0$ is not manifest at the classical level in the ambitwistor version reflects the fact that spaces of complex null geodesics may be constructed for any (globally hyperbolic) space--time, not just Einstein spaces, while the fact that it is non--chiral reflects the fact that, unlike twistor space, ambitwistor space has manifest parity invariance.

The genus zero NS--NS scattering amplitudes of this model were computed in \cite{Mason:2013sva}.  When $\Sigma\cong\CP^{1}$ there are three zero modes of both the $c$ and $\tilde c$ ghosts, and two zero modes of both $\gamma$ and $\tilde\gamma$.  At the level of a worldsheet correlation function, this means that the amplitude for $n$ external states is given by:
\be\label{gzamp}
\cM^{0}_{n}=\left\la c_{1}\tilde{c}_{1}U_1\;c_{2}\tilde{c}_{2}U_2\;c_{3}\tilde{c}_{3}V_{3}\;\prod_{i=4}^{n}\int_{\Sigma}\bar{\delta}(k_i\cdot P(z_i))\;V_i\right\ra,
\ee
which was evaluated in~\cite{Mason:2013sva} and shown to reproduce the Cachazo-He-Yuan representation of tree level graviton scattering amplitudes~\cite{Cachazo:2013hca}.  In this paper, we will generalize this computation to the torus, where internal states from the Ramond sector appear.

\subsection{Ramond sector vertex operators}
\label{sec:Rvertex}

Ramond sector vertex operators may also be constructed following the usual methods of string theory. We introduce spin fields $\Theta_{\alpha}$ of conformal weight $5/8$ for the worldsheet spinor $\Psi_{\mu}$, as well as the bosonization of the superconformal ghosts $\beta, \gamma$ \cite{Friedan:1985ey, Friedan:1985ge}.\footnote{We denote spinor indices as $\alpha, \beta$.  In ten space-time dimensions, these indices run from $\alpha=1,\ldots,32$.}  The latter involves a bosonic field $\phi$ with OPE
\begin{equation*}
	\phi(z)\;\phi(w)\sim -\ln|z-w|,
\end{equation*}
and insertions of the form $\e^{\ell \phi}$ have conformal weight $-\frac{\ell^2}{2}-\ell$.  Since a spin field has conformal weight $5/8$, the combination $\e^{-\phi/2}\Theta_{\alpha}$ takes values in $K_{\Sigma}$.  This allows us to define the fixed vertex operator
\be\label{fHfvo}
	c\tilde{c}\;\widehat{U}^{-1/2}=c\tilde{c}\,\e^{-\phi/2}\Theta_{\alpha}\zeta^{\alpha}\;\e^{\im k\cdot X}\;\tilde{U}\,,
\ee
of picture number $-1/2$, where $\tilde U$ may either be $\delta(\tilde\gamma) \tilde\epsilon\cdot\tilde\Psi$ if we wish to describe space--time gravitinos of polarization $\zeta^\alpha\tilde\epsilon_\mu$, or (in the heterotic model) an element $J$ of a worldsheet current algebra if we wish to describe gauginos of polarization $\zeta^\alpha$. In either case, $\tilde U$ also has holomorphic conformal weight 1 so that $\e^{-\phi/2}\Theta_\alpha \tilde U$ together balance the  holomorphic conformal weight $-2$ of $c\tilde c$.  The corresponding integrated vertex operator is
\be\label{fHvo}
	\int_\Sigma \bar\delta(k\cdot P)\,\widehat{V}^{-1} = \int_{\Sigma}\bar{\delta}(k\cdot P)\;\e^{-\phi/2}\Theta_{\alpha}\zeta^{\alpha}\;\e^{\im k\cdot X}\;\tilde{V} \, ,
\ee      
and has picture number $-1$. As in the NS sector, the scattering equation constraint $\bar\delta(k\cdot P)$ ensures that the expression under the integral sign is a (1,1)-form that may be integrated over $\Sigma$.

In type II ambitwistor strings, we may construct either R--NS or NS--R vertex operators, where the two sectors refer to the $\Psi$ and $\tilde\Psi$ systems, but where--in contrast to the usual string--both sectors are holomorphic. Imposing the standard GSO projections on each of these sectors implies that the spin fields must transform as a Weyl spinor in ten dimensions, leading to two gravitino states of either the same (IIB) or opposite (IIA) chiralities. There are also R--R sector $p$-form fields with vertex operators
\begin{align}\label{RR}
	c\tilde{c}\;\e^{-\phi/2}\Theta_\alpha\; \e^{-\tilde{\phi}/2}\tilde\Theta_\beta\;\gamma_{\alpha\beta}^{\mu_1\cdots\mu_p}\varepsilon_{\mu_1\cdots\mu_p}\; \e^{\im k\cdot X},
\end{align}
where as usual $p$ must be odd in the IIA model and even in the IIB. Thus the complete spectrum of Type IIA/B ambitwistor strings agrees with that of Type IIA/B supergravity in ten dimensions. Again we emphasize that triviality of the $XX$ OPE implies $\e^{\im k\cdot X}$ has vanishing conformal weight, so the ambitwistor string contains no massive states and no $\alpha'$ corrections.


\section{Amplitudes involving fermions}
\label{sec:fermions}

In this section, we compute the simplest amplitudes involving space--time fermions and verify them against known results.  For amplitudes involving more than four particles, we encouter the usual difficulties associated with space-time fermions in the RNS formalism.

\subsection{Three and four point amplitudes}
\label{sec:3pt}

The computation of three-- and four--point amplitudes involving two space--time fermions closely mirrors those of standard string theory (see {\it e.g.} \cite{Friedan:1985ge,Polchinski:1998rr}). In particular, for the three--point amplitude we use the correlation functions of the bosonized ghost and spin fields
\be
	\left\la \e^{-\phi_{1}/2}\,\e^{-\phi_{2}/2}\,\e^{-\phi_{3}}\right\ra=z_{12}^{-1/4}z_{23}^{-1/2}z_{31}^{-1/2} \qquad\qquad 
	\left\la \Theta_{1\alpha}\,\Theta_{2\beta}\,\psi^{\mu}_{3}\right\ra =\frac{\gamma^{\mu}_{\alpha\beta}}{z_{12}^{3/4}z_{23}^{1/2}z_{31}^{1/2}}\,,
\ee
where $\gamma^{\mu}_{\alpha\beta}$ are the $10$-dimensional gamma matrices. Individually, each of these correlators introduces branch cuts that cancel in the combined expression. There are no scattering equations to be imposed when $n=3$, so integrating out the $PX$ system just gives an overall momentum conserving $\delta$-function.

In the heterotic model, the correlator of three worldsheet currents $J$ gives 
\begin{equation*}
\left< J(z_1)J(z_2)J(z_3)\right> = \frac{\tr\left(\mathsf{T}^{a_1}\mathsf{T}^{a_2}\mathsf{T}^{a_3}\right)}{z_{12}z_{23}z_{31}},
\end{equation*}
and including the $c$ and $\tilde c$ ghost contributions leaves us with the heterotic amplitude
\be\label{h3pt2}
	\cM^{{\rm het}}(1_{f},2_{f},3_{b})
	=\delta^{10}\left(\sum_{i=1}^{3}k_i\right)\;\tr\left(\mathsf{T}^{a_1}\mathsf{T}^{a_2}\mathsf{T}^{a_3}\right)\;\zeta_{1}\cdot \slashed{\epsilon}_{3}\cdot\zeta_{2}\,,
\ee
the correct amplitude for two gauginos and a gauge boson. Likewise, the three-point amplitude for two gravitini and a graviton in the type II model is given by
\be\label{II3pt}
	\cM^{\mathrm{II}}(1_{f},2_{f},3_{b})
	=\delta^{10}\left(\sum_{i=1}^{3}k_i\right)\;\zeta_{1}\cdot \slashed{\epsilon}_{3}\cdot\zeta_{2}\ \tilde{\epsilon}_{1\mu}\tilde{\epsilon}_{2\nu}\tilde{\epsilon}_{3\rho} 
	T^{\mu\nu\rho},
\ee
where $T^{\mu\nu\rho}$ is built from the metric and momenta as
\be
 	T^{\mu\nu\rho}=\eta^{\mu\nu}(k_1-k_2)^{\rho}+\eta^{\nu\rho}(k_2-k_3)^{\mu}+\eta^{\rho\mu}(k_3-k_1)^{\nu}\,.
\ee
Thus~\eqref{II3pt} is the correct amplitude for the scattering of two gravitinos and one graviton.

\medskip

Four point amplitudes involving two bosons and two fermions may also be computed by following the usual steps in string theory. The main tool is the Ward identity~\cite{Knizhnik:1984nr}:
\be
	\left\la\Theta_{1\alpha}\Theta_{2\beta}\,\psi^{\mu}_{3}\,\psi^{\nu}_{4}\,\psi^{\rho a}_{4}\right\ra
	=\sum_{i\neq 4}\frac{M^{\nu\rho}_{i}}{z_{4i}}\left\la\Theta_{1\alpha}\,\Theta_{2\beta}\,\psi^{\mu}_{3}\right\ra\,,
\ee
where $M_i$ is a rotation matrix acting on the insertion at site $i$.  This allows us to compute the leading trace contribution to the four point amplitude in the heterotic model
\be\label{h4pt}
\begin{aligned}
	&\cM^{\rm het}(1_{f},2_{f},3_{b},4_{b}) =\left\la c_{1}\tilde{c}_{1}\widehat{U}^{-1/2}_{1}\;c_{2}\tilde{c}_{2}\widehat{U}^{-1/2}_{2}\;c_{3}\tilde{c}_{3}V^{-1}_{3}
	\int_{\Sigma}\bar\delta(k_4\cdot P)\,V_{4}\right\ra\\
	&\ =\delta^{10}\!\left(\sum_{i=1}^{4}k_{i}\right)\;\tr\left(\mathsf{T}^{a_1}\mathsf{T}^{a_2}\mathsf{T}^{a_3}\mathsf{T}^{a_4}\right)
	\int \bar{\delta}\!\left(\sum_{i=1}^{3}\frac{k_{4}\cdot k_{i}}{z_{4i}}\right)\frac{z_{31}}{z_{34}z_{41}} \times \\ 
	&\hspace{2cm}\left[\zeta_{1}\!\cdot\!\slashed{\epsilon}_{3}\!\cdot\!\zeta_{2}\sum_{j=1}^{3}\frac{\epsilon_{4}\cdot k_i}{z_{4i}}+\left(\frac{\zeta_{1}\!\cdot\!\gamma^{[\nu\sigma]}\!\cdot\!\slashed{\epsilon}_{3}\!\cdot\!\zeta_{2}}{z_{41}}-\frac{\zeta_{2}\!\cdot\!\gamma^{[\nu\sigma]}\!\cdot\!\slashed{\epsilon}_{3}\!\cdot\!\zeta_{1}}{z_{42}}\right)k_{4\;\nu}\epsilon_{4\;\sigma}\right. \\ 
	&\hspace{2.5cm}\left. \ +\ \frac{k_{4}\!\cdot\!\epsilon_{3}\ \zeta_{1}\!\cdot\!\slashed{\epsilon}_{4}\!\cdot\!\zeta_{2}-\epsilon_{4}\!\cdot\!\epsilon_{3} \ \zeta_{1}\!\cdot\!
	\slashed{k}_{4}\!\cdot\!\zeta_{2}}{z_{43}}\right] + \hbox{permutations} + \hbox{multi--trace}\,.
\end{aligned}
\ee
We have confirmed that upon reducing to four dimensional kinematics (always possible for four particle scattering) this expression produces the correct amplitude for the scattering of two gauginos and two gluons.  Similarly, one can compare \eqref{h4pt} with the single trace contribution to the $\alpha'\rightarrow 0$ limit of 4-point heterotic string amplitudes given in {\it e.g.}~\cite{Friedan:1985ge,Polchinski:1998rr}.  

A similar computation confirms that the type II amplitude
\be
\begin{aligned}\label{II4pt}
	&\cM^{\rm II}(1_{f},2_{f},3_{b},4_{b})=\left\la c_{1}\tilde{c}_{1}\widehat{U}^{-1/2,-1}_{1}\;c_{2}\tilde{c}_{2}\widehat{U}^{-1/2,-1}_{2}\;c_{3}\tilde{c}_{3} V^{-1,0}_{3}\;\int_{\Sigma}\bar\delta(k_4\cdot P)\, V_{4}\right\ra \\
	&\ =\delta^{10}\left(\sum_{i=1}^{4}k_{i}\right)\;\int \bar{\delta}\left(\sum_{i=1}^{3}\frac{k_{4}\cdot k_{i}}{z_{4i}}\right)z_{23}z_{31}\\ 
	&\hspace{2cm} \times\ \left[\zeta_{1}\cdot\slashed{\epsilon}_{3}\cdot\zeta_{2}\sum_{j=1}^{3}\frac{\epsilon_{4}\cdot k_i}{z_{4i}}+\left(\frac{\zeta_{1}\cdot\gamma^{[\nu\sigma]}\cdot\slashed{\epsilon}_{3}\cdot\zeta_{2}}{z_{41}}-\frac{\zeta_{2}\cdot\gamma^{[\nu\sigma]}\cdot\slashed{\epsilon}_{3}\cdot\zeta_{1}}{z_{42}}\right)k_{4\;\nu}\epsilon_{4\;\sigma}\right. \\ 
	&\hspace{3cm} \left. +\frac{k_{4}\cdot\epsilon_{3} \zeta_{1}\cdot\slashed{\epsilon}_{4}\cdot\zeta_{2}-\epsilon_{4}\cdot\epsilon_{3} \zeta_{1}\cdot\slashed{k}_{4}\cdot\zeta_{2}}{z_{43}}\right] \ \times \mathrm{Pf}\left(\widetilde{M}^{12}_{12}\right)
\end{aligned}
\ee
is the correct one for the scattering of two gravitinos and two gravitons.  Here, the $8\times 8$ matrix $\widetilde{M}$ is the same as the one appearing in the formul\ae\ of Cachazo {\it et al.}~\cite{Cachazo:2013hca} and accounts for the contribution to the amplitude from the NS $\tilde \Psi$ fields.

\subsection{Remarks on higher point amplitudes}
\label{sec:higher}

One of the attractive features of the $n$-particle formul\ae\ of~\cite{Cachazo:2013hca} is their compactness.  It is natural to ask if we can find similarly compact expressions for scattering amplitudes involving two gravitinos and an arbitrary number of gravitons.  It is easy to see that in the type II ambitwistor string, these amplitudes are determined by the correlator
\begin{multline*}
\left\la c_{1}\tilde{c}_{1}\widehat{U}^{-1/2,-1}_{1}\;c_{2}\tilde{c}_{2}\widehat{U}^{-1/2,-1}_{2}\;c_{3}\tilde{c}_{3} V^{-1,0}_{3}\;\prod_{i=4}^{n}\int_{\Sigma}\bar{\delta}(k_{i}\cdot P)\;V_{i}\right\ra \\
= \int\prod_{j=4}^{n}\bar{\delta}\left(\sum_{k\neq j}\frac{k_{j}\cdot k_{k}}{z_{jk}}\right) z_{23}z_{31}\; \mathrm{Pf}\left(\widetilde{M}^{12}_{12}\right) \\
\times \left\la\Theta_{1}\cdot\zeta_{1}\;\Theta_{2}\cdot\zeta_{2}\;\epsilon_{3}\cdot\psi_{3}\prod_{j=4}^{n}(\epsilon_{j}\cdot P_{j}+\epsilon_{j}\cdot\psi_{j}\;k_{j}\cdot\psi_{j})\e^{\im\sum k\cdot X}\right\ra.
\end{multline*}
involving one copy of the Pfaffian from the NS sector. Unfortunately, due to the non--polynomial nature of the spin field $\Theta_\alpha$, we have been unable to find a compact, closed--form expression for this correlator. This is as expected: the RNS formulation of a string theory (including an ambitwistor string) obscures space--time supersymmetry and makes calculations of scattering amplitudes involving arbitrary numbers of fermions rather laborious.

Berkovits recently constructed a pure spinor version of the ambitwistor string~\cite{Berkovits:2013xba}, which possesses manifest space--time supersymmetry and so may be expected to be able to treat scattering amplitudes more easily by adapting the techniques of~\cite{Mafra:2011nv}.  Indeed, it has now been shown that the genus zero amplitudes of this model correspond to those of ten dimensional $\cN=1$ super-Yang-Mills in the heterotic case, and type II supergravity in the type II case~\cite{Gomez:2013wza}. We note however that the Pfaffians appearing in~\cite{Cachazo:2013hca,Cachazo:2013iea} for the NS--NS sector seem inevitably to point to an RNS model with real worldsheet spinors.

In four dimensions, compact expressions for all tree amplitudes, of any helicity type, in $\cN=8$ supergravity are available using the twistor string~\cite{Cachazo:2012kg,Skinner:2013xp}. The integrals over the moduli space occurring in these representations are also frozen in terms of the external momenta, which are now manifestly on--shell. The constraints imposed on the twistor string also imply that the scattering equations are satisfied~\cite{Witten:2004cp}.


\section{Ambitwistor strings at genus one}
\label{sec:1loop}

We now investigate the ambitwistor string a genus one. We begin by considering the scattering equations on an $n$-pointed elliptic curve, showing that two different types of equation must be imposed: on the residues of $P^2$ and on $P^2$ itself. We then proceed to study the partition function and worldsheet correlation functions.

\subsection{The scattering equations at genus 1}
\label{sec:seqs}

At genus one, the moduli of the Riemann surface includes the complex structure parameter $\tau$ of the unmarked curve, as well as the markings (up to the freedom to fix one marked point). So after converting $n-1$ of the vertex operators to integrated vertex operators, the measure on the moduli space also involves an insertion
\be
	\int_\Sigma \tilde b\, \mu\ \times\  \bar\delta\!\left(\int_\Sigma P^2\mu\right)
	=\tilde b_0 \, \bar\delta(P^2(z_0;\tau))\,,
\label{measurefactor}
\ee
where $\mu$ is the Beltrami differential associated to changes in the complex structure of the elliptic curve. As usual, the insertion of $\tilde b_0$ serves to absorb the single constant zero mode of $\tilde b$ at genus 1, so its insertion point is arbitrary. The remaining $\delta$-function forms part of the genus 1 scattering equations. It should be interpreted as
\be	
	\bar\delta(P^2(z_0;\tau)) = \rd\bar\tau \frac{\del}{\del\bar\tau} \left(\frac{1}{P^2(z_0;\tau)}\right) \, ,
\ee
and fixes the integral over $\tau$. Thus, in addition to imposing ${\rm Res}_i P^2(z_i) = 0$ (as at $g=0$ but now for $n\!-\!1$ of the marked points)  we also impose that $P^2$ itself vanishes at some other point $z_0$. To understand these two different types of scattering equations, note first that (as we will find below) $P^2$ has at most simple poles at the marked points and no other singularities. Thus, once we impose that the residue of $P^2$ vanishes at $n\!-\!1$ of these marked points we know that $P^2$ must in fact be globally holomorphic over $\Sigma$, since the remaining residue must inevitably vanish. However, at genus one there exists a unique globally holomorphic quadratic differential which must be constant, since $K^2_\Sigma \cong\cO$. The final scattering equation $P^2(z_0)=0$ ensures that this constant piece also vanishes. 

Altogether then, the $n$ scattering equations
\be
	{\rm Res}_i\, P^2 = 0 \quad i=2,\ldots,n \qquad\hbox{and}\qquad P^2(z_0)=0
\ee
are exactly what is needed to ensure that $P^2$ in fact vanishes everywhere on the worldsheet. This vanishing is the content of the scattering equations at any genus, and is crucial to ensure that the gauge symmetry $\delta X^\mu = \tilde c P^\mu$ is consistent in the presence of vertex operators. It is precisely because the scattering equations hold that we must interpret the target space of the string theory as ambitwistor space, not $T^*M$.

\medskip

We can write the genus 1 scattering equations in a more explicit form by performing the $XP$ path integral.  As in the genus zero case discussed in~\cite{Mason:2013sva}, we do this by treating the plane wave $\e^{\im k_i\cdot X(z_i)}$ factors in the vertex operators as localized contributions to the worldsheet action. At any genus, zero modes of $X^\mu$ must be constant, and integrating over these constants leads to the ten--dimensional momentum conserving $\delta$-function $\delta^{10}\!\left(\sum_i k_i\right)$. The path integral over the non--zero modes of $X$  imposes the constraint 
\be
	\delbar P_\mu(z) = 2\pi\im\, {\rm d}z\wedge{\rm d}\bar z\;\sum_{i=1}^n k_{i\mu} \,\delta^2(z-z_i) 
\label{Pconstr}
\ee
saying that $P_\mu$ is holomorphic except for poles at the vertex operators. However, unlike for a Riemann sphere, an elliptic curve possesses a globally holomorphic abelian differential that, using the identification $E_\tau\cong \C/\Lambda$, may be written as the holomorphic 1-form $\rd z$ on the complex plane.  Thus at genus one~\eqref{Pconstr} has a homogeneous solution $P_\mu(z)=p_\mu \rd z$ where $p_\mu$ are constants. These constants are the zero modes of $P_\mu$ and must be separately integrated over. 
Accounting for the poles, the general solution of~\eqref{Pconstr} is
\be
	P_\mu(z) = p_\mu \rd z + \sum_{i=1}^n k_{i\mu}\tilde S_1(z,z_i;\tau)\,,
\label{Psolv}
\ee
where 	
\be	
	\tilde S_1(z,z_i;\tau)= \left(\frac{\theta_{1}'(z-z_i,\tau)}{\theta_{1}(z-z_i,\tau)} + 4\pi\frac{{\rm Im}(z-z_i)}{{\rm Im}(\tau)}\right)\rd z
\label{tildeS1def}
\ee
is the propagator for the $PX$-system on an elliptic curve. Note that 
\be
	\tilde S_1(z,z_i;\tau) = \rd z \frac{\del}{\del z}  G(z,z_i;\tau)
\ee
where 
\be
	G(z,z_i;\tau)=-\ln |E(z,z_i;\tau)|^2 + {2\pi}\frac{({\rm Im}(z-z_i))^2}{{\rm Im}(\tau)}
\label{nonchiralprop}
\ee
is the usual genus one propagator for a non--chiral scalar, written in terms of the prime form $E(z,w;\tau)$.

The term proportional to ${\rm Im}(z-z_i)$ in~\eqref{tildeS1def} ensures that $\tilde S_1$ is orthogonal to the zero mode $P_\mu(z)=p_\mu\rd z$. However, on the support of the momentum conserving $\delta$-function, the sum $\sum_{i=1}^n k_{i\mu}\tilde S_1(z,z_i)$ in~\eqref{Psolv} is independent of ${\rm Im}(z)$, so that~\eqref{Psolv} is meromorphic in $z$ as required. The dependence of the sum on the ${\rm Im}(z_i)$ can be absorbed into a shift of the zero mode $p_\mu$ if need be. However, it is simpler to treat this term as part of $\tilde S_1$ as it ensures that~\eqref{tildeS1def} behaves as
\be
	\tilde S_1(z,z_i;\tau) = \tilde S_1(z,z_i;\tau+1) = \tilde S_1\left(\frac{z}{\tau},\frac{z_i}{\tau}\,;-\frac{1}{\tau}\right)\,,
\ee
under modular transformations, where we recall that $\tilde S_1(z,z_i)$ is a (1,0)-form in $z$ and a scalar in $z_i$. 

Using~\eqref{Psolv} and the fact that $k_i^2=0$, the $\bar\delta$-functions in the integrated vertex operators now impose the constraint
\be
	0=k_i\cdot p + \sum_{j\neq i} k_i\cdot k_j\, \tilde S_1(z_i,z_j;\tau)
\label{vertexseq}
\ee
at all but one of the marked points. The remaining constraint comes from $\bar\delta(P^2)$ in the measure for integrating over the moduli space. This imposes
\be
	0=p^2(\rd z)^2 + (\rd z) \sum_{i=1}^n p\cdot k_i\,\tilde S_1(z,z_i;\tau) + \sum_{i\neq j} k_i\cdot k_j\,\tilde S_1(z,z_i;\tau)\tilde S_1(z,z_j;\tau) \, ,
\label{moduliseq}
\ee
where the second sum runs over both $i$ and $j$. Equations~\eqref{vertexseq}-\eqref{moduliseq} are the genus one analogue of the genus zero scattering equations used in~\cite{Cachazo:2013hca,Cachazo:2013iea,Mason:2013sva}. It would be interesting to compare them to the genus one saddle point equations found by Gross and Mende~\cite{Gross:1987ar}, although we note that the scattering equations here depend on the zero mode $p_\mu$ of the field $P_\mu$ that is absent in usual, second--order formulations of string theory.

\medskip

The $n$ scattering equations completely fix the integral over the $n$-dimensional moduli space $\cM_{1,n}$ of $n$-pointed genus 1 curves in terms of the external momentum $k_i$ and the zero mode $p$. However, the zero modes $p$ of $P(z)$ are not fixed. These variables are just the usual (generically off--shell) momentum circulating around the loop in the corresponding 1--loop Feynman diagrams. Thus, in contrast to standard non--chiral string theory, the ambitwistor string explicitly introduces a loop momentum and, if one wishes to evaluate the full amplitude rather than merely the loop integrand, the loop integral $\rd^{10}p$ must be performed explicitly after evaluating the worldsheet correlation functions (for which see sections~\ref{sec:even}--\ref{sec:odd}).

The fact that the loop momentum appears explicitly in this formalism has a very important consequence. Usual string theory is UV finite because (given a well--defined worldsheet CFT) its bosonic moduli space is essentially $\overline{\cM}_{g,n}$ -- the Deligne--Mumford moduli space of marked curves\footnote{The statement that the integral is over the moduli space $\overline{\cM}_{n,g}$ rather than over the non--compact Teichm{\"u}ller space makes crucial use of invariance under the modular group Sp$(2g,\mathbb{Z})$. The full bosonic moduli space also includes the space of worldsheet instantons over each point of $\overline{\cM}_{g,n}$. In flat space--time $\R^{1,9}$ these are just the (constant) zero modes of $X^\mu(z,\bar z)$ and, in the presence of vertex operators, the corresponding integral yields a momentum conserving $\delta$-function. In compactifications the worldsheet instanton moduli space can be more complicated, but still leads to no new divergences essentially because it is either compact or admits a natural compactification.}. Worldsheet correlation functions have singularities on the boundary $\overline{\cM}_{g,n}\backslash {\cM}_{g,n}$ of this space, but these correspond to (physically important) IR divergences. See {\it e.g.}~\cite{Witten:2012bh,Witten:2013tpa} for a recent comprehensive discussion.

By contrast, in the case of ambitwistor strings the moduli space also includes an integral over a copy of real (Minkowskian) momentum space $\R^{1,9}\subset\C^{10}$ corresponding to the $P_\mu$ zero modes. This space is non--compact, and this final integral is potentially divergent. This is how the chiral ambitwistor string can be both a string theory and yet be equivalent to a pure (massless) supergravity in the target space -- potential UV divergences come not from the integral over the (compact) moduli space of marked curves, but from the non--compact loop integrals over zero--modes of $P(z)$. In particular, we expect type II supergravities to display a quadratic\footnote{In dimensional regularization, such power law divergences are absent, so 10d supergravity will be accidentally finite until two loops} UV divergence at one loop in ten dimensions. It would be interesting to see this behaviour in the final expressions for worldsheet correlation functions below.

At genus $g$ we expect to have a total of $n+3g-3$ scattering equations, of which (for $g\geq2$) $n$ would be of the type $k_i\cdot P(z_i)=0$ constraining the residues of $P(z)$ to vanish at the vertex operators, while $3g-3$ would be of the type $P^2(z_r)=0$ constraining the possible holomorphic quadratic differential $P^2$ to vanish at $3g-3$ points $z_r\in \Sigma$. Since $h^0(\Sigma,K^2(z_1+\cdots+z_n))= n+3g-3$ these scattering equations suffice to impose $P^2(z)=0$ globally over the marked Riemann surface, ensuring as in~\cite{Mason:2013sva} that the true target space of the string is ambitwistor space $\PA$. On the other hand, there are $g$ holomorphic Abelian differentials $\omega_a$ (with $a=1,\ldots,g$), these higher genus amplitudes will involve an integral over the zero modes $\prod_{a} \rd^{10}p_a$ of $P(z)$, corresponding to the loop momenta at $g$ loops in field theory. Again, we expect these integrals to diverge, both in the UV and IR.


\subsection{Modular invariance and the partition function}
\label{sec:modular}

As usual in string theory, the path integrals over the non--zero modes of the fields are non--trivial at genus one, even in the absence of any vertex operator insertions. For the odd spin structure, the $\Psi^\mu$ and $\tilde\Psi^\mu$ fields each have (constant) zero modes which, in the absence of vertex operator insertions, kill the contribution of the odd spin structure to the partition function. For an even spin structure, neither the fermionic fields $\Psi$, $\tilde\Psi$ nor the associated $\beta\gamma$ and $\tilde\beta\tilde\gamma$ ghost systems have any zero modes. Therefore the partition function becomes
\be
	Z_{\bd\alpha}(\tau)\tilde Z_{\bd\beta}(\tau) =	\frac{{\det'(\delbar_{\,T_\Sigma})}^2}{{\det'(\delbar_{\,\mathcal{O}})}^{10}}
	\frac{{\rm Pf}(\delbar_{K_\Sigma^{1/2}(\bd\alpha)})^{10}}{\det(\delbar_{T_\Sigma^{1/2}(\bd\alpha)})}
	\frac{{\rm Pf}(\delbar_{K_\Sigma^{1/2}(\bd\beta)})^{10}}{\det(\delbar_{T_\Sigma^{1/2}(\bd\beta)})}
	= \frac{1}{\eta(\tau)^{16}}\frac{\theta_{\bd{\alpha}}(0;\tau)^4}{\eta(\tau)^4}\frac{\theta_{\bd{\beta}}(0;\tau)^4}{\eta(\tau)^4} \,,
\label{partition}
\ee
where $\bd\alpha$ and $\bd\beta$ are the spin structures associated to $\{\Psi,\gamma,\beta\}$ and $\{\tilde\Psi,\tilde\gamma,\tilde\beta\}$ respectively, and $\eta(\tau)$ is the Dedekind eta function.

We can combine these partition functions to form modular invariants. To begin with, the standard GSO projections of Type II strings correspond to the $g=1$ partition functions 
\be
\begin{aligned}
	Z_{\rm IIA}(\tau) &= \left(Z_1+\sum_{\bd\alpha=2,3,4}(-1)^{\bd\alpha}Z_{\bd\alpha}\right)\left(\tilde Z_1-\sum_{\bd\alpha=2,3,4}(-1)^{\bd\alpha}\tilde Z_{\bd\alpha}\right)\\
	Z_{\rm IIB}(\tau) &= \left(Z_1+\sum_{\bd\alpha=2,3,4}(-1)^{\bd\alpha}Z_{\bd\alpha}\right)\left(\tilde Z_1+\sum_{\bd\alpha=2,3,4}(-1)^{\bd\alpha}\tilde Z_{\bd\alpha}\right)\,,
\end{aligned}
\label{typeII}
\ee
for type IIA and type IIB ambitwistor strings. Here $Z_1$ and $\tilde Z_1$ are the (vanishing) partition functions of the $\Psi$ and $\tilde\Psi$ systems in the odd spin structure. As usual, both these partition functions vanish as a consequence of the Jacobi `abstruse identity' ${\theta_2(\tau)}^4-{\theta_3(\tau)}^4+{\theta_4(\tau)}^4=0$ that  reflects space--time supersymmetry and imposes the one--loop vanishing of the space--time cosmological constant. 

We can also construct a type 0 ambitwistor string by requiring the $\Psi$ and $\tilde\Psi$ systems to have the {\it same} spin structures. This choice breaks space--time supersymmetry. However, unlike the real partition function $\propto |\theta_2(\tau)|^N + |\theta_3(\tau)|^N + |\theta_4(\tau)|^N$ of non--chiral Type 0 strings which is modular for any value of $N$, the chiral partition function $\propto {\theta_2(\tau)}^8 + {\theta_3(\tau)}^8 + {\theta_4(\tau)}^8$ of the type 0 ambitwistor string can be modular only in $8k+2$ space-time dimensions. 

\medskip

The above partition functions~\eqref{typeII} are modular functions of weight $-8$. Including the integral over the zero modes of $X$ and $P$, together with the zero modes of the $bc$ and $\tilde b\tilde c$ ghost systems, the full genus one partition function of the type II string is formally
\be
	\cZ_{\rm{IIA/B}} = \int \frac{\rd^{10}x\,\rd^{10}p}{({\rm vol}\, \C^*)^2} \ \bar\delta\!\left(p^2(\rd z)^2\right)\,Z_{\rm II A/B}(\tau)\, \rd\tau \,,
\label{Z}
\ee
where we have solved~\eqref{Pconstr} to find $P_\mu(z) = p_\mu\rd z$ in the absence of any vertex operator insertions. Here $x^\mu$ is just a constant zero mode of $X^\mu$, while $p_\mu$ is the coefficient of the abelian differential $\rd z$ arising as a zero mode of $P_\mu$. Under the modular transformation $\tau\to-1/\tau$ this differential behaves as $\rd z \to \rd z/\tau$, so we should also transform 
\be
	p_\mu \to \tau p_\mu
\ee
to ensure that the zero mode $p_\mu\rd z$ itself is invariant. With this definition, the loop integral measure $\rd^{10}p$ acquires a factor of $\tau^{10}$ under this modular transformation. This compensates the weight of the modular function $Z_{\rm IIA/B}(\tau)\,\rd\tau$ so that~\eqref{Z} is invariant.

The factor of $1/({\rm vol\,}\C^*)^2$ arises from fixing the zero modes of the $c$ and $\tilde c$ ghosts. The $c$ ghost zero mode may be used to fix the insertion point of $\bar\delta(p^2(\rd z)^2)$ to any point on the torus. Recalling that the $\tilde c$ ghost is associated to the transformation $\delta X^\mu = \tilde c P^\mu$ that allowed us to translated $X$ along the null geodesic in the direction of $P$, we may use the remaining ${\rm vol}\,\C^*$ factor to fix one of the $x$ integrals, picking a representative point on each null geodesic. Combining this action with the constraint $p^2=0$ we see that the integral over zero modes of $X$ and $P$ is really an integral over the target space $\PA$. This is as expected in string theory, and once again emphasizes the fact that the target space of this chiral model is best thought of as ambitwistor space, rather than space--time.


\subsection{NS--NS scattering amplitudes at genus 1}
\label{sec:1loopamps}

We now wish to consider the contribution to the $n$--point scattering amplitude of particles in the NS--NS sector of ten dimensional supergravity --- {\it i.e.}, gravitons, $B$--fields and dilatons --- from the genus one ambitwistor string. As in section~\ref{sec:review}, for momentum eigenstates these particles may be described either by fixed vertex operators
\be
	c\tilde c U_i(z_i) = c\tilde c\,\delta(\gamma)\,\delta(\tilde\gamma)\, \epsilon_i \cdot\Psi(z_i)\,\tilde\epsilon_i\cdot\tilde\Psi(z_i)\,\e^{\im k_i\cdot X(z_i)}
\label{fixedU}
\ee
or by the corresponding integrated vertex operators 
\be
\begin{aligned}
	\int_\Sigma\bar\delta\!\left(k_i\cdot P(z_i)\right)V_i(z_i) &\ \ = \\
	&\hspace{-2.5cm}\int_\Sigma\bar\delta\!\left(k_i\cdot P(z_i)\right)\!
	\left[\phantom{\tilde\Psi}\hspace{-0.3cm}\epsilon_i\cdot P + \epsilon_i\cdot \Psi\, k_i\cdot\Psi\right]
	\left[\tilde\epsilon_i\cdot P + \tilde\epsilon_i\cdot \tilde\Psi\, k_i\cdot\tilde\Psi\right](z_i)\,\e^{\im k_i\cdot X(z_i)}
\end{aligned}
\ee
that follow from~\eqref{fixedU} via the descent procedure. We will consider the case that the fermions have even or odd spin structures separately.

\subsubsection{Even spin structure}
\label{sec:even}

In an even spin structure, neither the worldsheet fermions $\Psi^\mu$, $\tilde\Psi^\mu$ nor the ghosts $\gamma$, $\tilde\gamma$ have zero modes, so we want no $U$ insertions. However, the $c$ and $\tilde c$ ghosts have one zero mode each, corresponding to constant translations around the torus, or along the null geodesic $x^\mu(\lambda) = x^\mu + \lambda p^\mu$. This freedom is fixed by one insertion of $c\tilde c V$. We thus wish to compute
\be
	\mathcal{M}_n^{1;\,{\rm even}} = \left< b_0 \tilde b_0\,\bar\delta(P^2)\, c\tilde cV_1(z_1) \prod_{i=2}^n \int \bar\delta(k_i\cdot P(z_i))V_i(z_i)\right\rangle \, ,
\label{1loop}
\ee
where the factor of $\bar\delta(P^2)$ in the measure was explained above.

Because none of the vertex operators involve $\delta(\gamma)$ or $\delta(\tilde\gamma)$, the correlator of the $\Psi$ fields and of the $\tilde\Psi$ fields each lead to Pfaffians of $2n\times 2n$ matrices $M'_{\bd\alpha}$ and $\widetilde M'_{\bd\beta}$. In other words, unlike at genus zero~\cite{Cachazo:2013hca,Cachazo:2013iea,Mason:2013sva}, no rows or columns are removed from these matrices. The matrix $M'_{\bd\alpha}$ has elements
\be
	M'_{\bd\alpha} = \begin{pmatrix} 
			A & -{C'}^{\rm T}\\
			C' & B
		\end{pmatrix}
\ee
where 
\be
	A_{ij} = k_i\cdot k_j\, S_{\bd\alpha}(z_{ij};\tau) 	\qquad\qquad B_{ij} = \epsilon_i\cdot\epsilon_j \,S_{\bd\alpha}(z_{ij};\tau)
	\qquad\qquad C'_{ij} = \epsilon_i\cdot k_j\, S_{\bd\alpha}(z_{ij};\tau)
\label{ABCg=1}
\ee
and $A_{ii}= B_{ii}=C'_{ii}=0$ again on account of $\epsilon_i\cdot k_i = k_i^2 = 0$. In this matrix, 
\be
	S_{\bd \alpha}(z_{ij},\tau) = \frac{\theta_{1}'(0;\tau)}{\theta_{1}(z_{ij};\tau)}\frac{\theta_{\bd \alpha}(z_{ij};\tau)}{\theta_{\bd \alpha}(0;\tau)}\sqrt{\rd z_i}\sqrt{\rd z_j}
\label{Szego}
\ee
is the $g=1$ free fermion propagator, or Szego kernel, in the even spin structure $\bd{\alpha}$. We have defined this to be a half--form in both $z_i$ and $z_j$ (like $\Psi(z_i)\Psi(z_j)$) so that  under a modular transformation it simply changes to a Szego kernel in a different (even) spin structure ({\it i.e.} it does not acquire any factors of $\sqrt\tau$).

The elements of $M'_{\bd\alpha}$ arise from considering contractions between the various $\Psi$ insertions at points $z_i$ and $z_j$ (with $i\neq j$) on the worldsheet, where we recall that the $i^{\rm th}$ vertex operator involves a term $\epsilon_i\cdot\Psi(z_i)\,k_i\cdot\Psi(z_i)$. As at genus zero~\cite{Skinner:2013xp,Mason:2013sva}, we may incorporate the contributions from the $\epsilon_i\cdot P(z_i)$ factors in the vertex operators by modifying the matrix $C'\to C$, where the off--diagonal elements are unchanged, but where the diagonal elements now become\footnote{Recall that $\tilde S_1(z,w;\tau)$ is a (1,0)-form in $z$ and a scalar in $w$.}
\be
	C_{ii} = \epsilon_i\cdot p\ \rd z_i + \sum_{j\neq i} \epsilon_i\cdot k_j\, \tilde S_1(z_i,z_j;\tau) \,,
\label{Cdiag}
\ee
independent of the spin structure. That is, the diagonal elements $C_{ii} = \epsilon_i\cdot P(z_i)$, with $P(z)$ given by~\eqref{Pconstr} and we use $\epsilon_i\cdot k_i=0$ before taking the $z\to z_i$ limit. Alternatively, normal ordering of the vertex operators means that we should ignore the divergent contribution obtained if one sets $z\to z_i$ in~\eqref{Pconstr} before contracting with the polarization tensor $\epsilon_i$. Thus, in an even spin structure $\bd\alpha$, the vertex operators contribute a factor of ${\rm Pf}(M_{\bd\alpha})\,{\rm Pf}(\widetilde M_{\bd\beta})$ to the string correlation function, where 
 \be
 	M_{\bd\alpha}= \begin{pmatrix}
		A & -{C}^{\rm T}\\
			C & B
		\end{pmatrix}
\ee
and $\tilde M_{\bd\beta}$ is similar but with tilded polarization tensors and a (perhaps) different spin structure $\bd\beta$. On the support of the scattering equations, these Pfaffians are invariant under the target space gauge transformations $\epsilon_i \to \epsilon_i+k_i$, as follows from worldsheet BRST invariance.

Combining this with the non--trivial path integral that gave the partition function and summing over even spin structures $\bd\alpha$ and $\bd\beta$ according to the type II GSO projection, we obtain
\be
\begin{aligned}
	\cM_n^{1;\,{\rm even}}&= \delta^{10}\!\left(\sum_{i=1}^n k_i\right)\int 
	\rd^{10}p\wedge\rd\tau \ \bar\delta\!\left(P^2(z_1;\tau)\right)\,\prod_{j=2}^n \bar\delta(k_j\cdot P(z_j)) \\
	&\hspace{4cm} \times\ \sum_{\bd\alpha;\bd\beta} (-1)^{\bd\alpha+\bd\beta} Z_{\bd\alpha;\bd\beta}(\tau)
	\ {\rm Pf}(M_{\bd\alpha})\,{\rm Pf}(\widetilde{M}_{\bd\beta})
\end{aligned}
\label{evenamp}
\ee
as the contribution to 1--loop scattering amplitudes from even spin structures. Note that the integrand in~\eqref{evenamp} is a (top,top) form on $\cM_{n,1}$; the product of the two Pfaffians transforms as a quadratic differential at each marked point $z_i$ for $i\in\{1,\ldots,n\}$, while the constraints $\prod_{j=2}^n\,\bar\delta(k_j\cdot P(z_j))$ provide holomorphic conformal weight $-1$ at all the marked points except $z_1$, whereas the constraint $\bar\delta\!\left(P^2(z_1;\tau)\right)$ provides holomorphic weight $-2$ at $z_1$.

As mentioned above, these scattering equation constraints fix the vertex operator insertion points $z_i$ and the worldsheet complex structure $\tau$ in terms of the external and loop momenta $k_i$ and $p$. The integral over the loop momentum $p$ must be treated as a contour integral and is expected to diverge on the physical contour $\R^{9,1}\subset \C^{10}$. Notice that the loop momentum appears in the Pfaffians, through the diagonal elements~\eqref{Cdiag} of $C$, as well as in the scattering equations. Modular invariance of the right hand side of~\eqref{evenamp} follows trivially from the modular invariance of the partition function; indeed, we included form weights in the elements of $M_{\bd\alpha}$ and $\widetilde M_{\bd\beta}$ precisely to ensure that their Pfaffians are invariant under modular transformations, up to a change in spin structure.

\subsubsection{Odd spin structure}
\label{sec:odd}

At genus one, there is a single odd spin structure corresponding to periodic boundary conditions around each of the two non-trivial cycles on the torus.  In this spin structure the the ghosts and antighost have one, constant zero mode each.  The zero modes of the antighosts correspond to fermionic moduli, which as in the RNS string we fix by inserting two picture changing operators
\begin{align}\label{PCO}
	 \Upsilon_0=\bar\delta(\beta)(P\cdot\Psi+\tilde b\gamma) \qquad\qquad \widetilde{\Upsilon}_0=\bar\delta(\tilde\beta)(P\cdot\tilde\Psi+\tilde b\tilde\gamma)\,.
\end{align}
At least at genus one, there are no spurious singularities and BRST invariance ensures the amplitude is independent of the choice of insertion point of these operators.

Since each component of the fermionic fields $\Psi^\mu$ and $\tilde\Psi^\mu$ also has a zero mode, as usual only amplitudes with at least five particles receive any contributions from this spin structure. For $n\geq5$ the amplitude receives a contribution from the worldsheet correlator
\begin{align}\label{oddamp}
 	{\cM}^{1;\;\rm odd}_n
 	=\left<b_0\tilde b_0\,\bar\delta(P^2(z_0))\,\Upsilon_{0}\widetilde{\Upsilon}_{0}\ c_1\tilde c_1U(z_1) \prod_{i=2}^n\int\bar\delta(k_i\cdot P(z_i))V(z_i)\right>.
\end{align}
Evaluating this correlator leads again to Pfaffians of  $2n\times2n$ matrices. For the $\Psi$ system we obtain the matrix
\begin{align}\label{oddM}
 M=\begin{pmatrix}
 	A & -C^{\rm T}\\
	C & B
	\end{pmatrix}\,,
\end{align}
where the entries now depend on the $\Psi$ zero modes $\Psi_0$. For $i\neq j$ we have
\be\label{oddentries}
\begin{aligned}
	A_{ij}&=k_i\cdot k_j\;S_1(z_{ij};\tau)+k_i\!\cdot\!\Psi_0\ k_j\!\cdot\!\Psi_0\qquad &i,j\neq 1\\
	B_{ij}&=\epsilon_i\cdot\epsilon_j\;S_1(z_{ij};\tau)+\epsilon_i\!\cdot\!\Psi_0\ \epsilon_j\!\cdot\!\Psi_0 &\\
	C_{ij}&=\epsilon_i\cdot k_j\;S_1(z_{ij};\tau)+\epsilon_i\!\cdot\!\Psi_0\ k_j\!\cdot\!\Psi_0\,, &
\end{aligned}
\ee
whenever $i\neq1$, and diagonal entries
\be\label{odddiagC}
\begin{aligned}
 	C_{ii}=-\epsilon_i\!\cdot\! P(z_0)\rd z_i-\sum_{j\neq i}^n\epsilon_i\!\cdot\! k_j\ S(z_{ij};\tau)\,,
\end{aligned}
\ee
again for $i\neq1$. When $i=1$, the entries of $A$ and $C$ are modified to
\be\label{odd1entries}
\begin{aligned}
	A_{1j}&= P(z_0)\!\cdot\! k_j\ S_1(z_{0j})+P(z_0)\!\cdot\!\Psi_0\ k_j\!\cdot\!\Psi_0\\
 	C_{11}&=\epsilon_i\!\cdot\! P(z_0)\ S_1(z_{10})+\epsilon_i\!\cdot\!\Psi_0\ P(z_0)\!\cdot\!\Psi_0\, ,
\end{aligned}
\ee
as they originate from contractions involving the picture changing operator. In these expressions, $S_1(z_{ij};\tau)$ is the free fermion propagator
\be\label{oddprop}
	S_1(z_{ij};\tau):=\left(\frac{\theta'_1(z_i-z_j;\tau)}{\theta_{1}(z_i-z_j;\tau)}+4\pi\frac{{\rm Im}(z_i-z_j)}{{\rm Im}(\tau)}\right)\,\sqrt{\rd z_i}\sqrt{\rd z_j}
\ee
orthogonal to the zero mode. Again we have chosen to treat this as a half--form in each of $z_i$ and $z_j$, making it invariant under modular transformations. Note also that the zero mode $\Psi_0^\mu = \Psi_{0z}^\mu\sqrt{\rd z}$, where $\Psi^\mu_{0z}$ are anticommuting constants. 

After performing all contractions to obtain the Pfaffian of $M$ (and a Pfaffian of a similar matrix $\widetilde M$), we must still perform the path integral over all the fields. Here we find simply
\begin{align}
	\frac{{\det'(\delbar_{\,T_\Sigma})}^2}{{\det'(\delbar_{\,\mathcal{O}})}^{10}}
	\frac{{\rm Pf}(\delbar_{K_\Sigma^{1/2}})^{10}}{\det(\delbar_{T_\Sigma^{1/2}})}
	\frac{{\rm Pf}(\delbar_{K_\Sigma^{1/2}})^{10}}{\det(\delbar_{T_\Sigma^{1/2}})}=1\,,
\end{align}
where we have used the fact that $K_\Sigma^{1/2} = T_\Sigma^{1/2}=\cO$ for the odd spin structure.

Finally then, including the integration over zero modes, the contribution of the odd spin structure to $n\geq5$ particle amplitudes is
\begin{multline}\label{oddamp2}
 	\cM^{1;\;\rm odd}_n=\delta^{10}\left(\sum k_i\right)\int \rd^{10}p\, \rd^{10}\Psi_{0}\, \rd^{10}\tilde\Psi_{0}\,\rd\tau\,\bar\delta(P^2(z_1))\prod_{i=2}^n\bar\delta(k_i\cdot P(z_i)) \\
\times {\rm Pf}(M)\; {\rm Pf}(\tilde M)\;\frac{\rd z_1}{(\rd z_0)^3},
\end{multline}
where $\rd^{10}\Psi_0$ and $\rd^{10}\tilde{\Psi}_{0}$ are the integrals over the $\Psi$ and $\tilde\Psi$ zero modes, while the ratio $\rd z_1/(\rd z_0)^3$ arises from the zero modes of the ghost and antighosts in the picture changing operators. It is easy to see that \eqref{oddamp2} is invariant under $\tau\rightarrow\tau +1$. Under $\tau\rightarrow -1/\tau$, invariance of $p\rd z$ again implies that $\rd^{10}p\rightarrow\tau^{10}\,\rd^{10}p$.  Likewise, invariance of $\Psi_0\sqrt{\rd z}$ implies that the Berezinian integration 
$\rd^{10}\Psi_{0}\rightarrow\tau^{-5}\,\rd^{10}\Psi_{0}$, and similarly for the $\tilde\Psi$ zero modes.  Therefore, under $\tau\to-1/\tau$,
\be
 	\rd^{10}p\,\rd^{10}\Psi_0\,\rd^{10}\tilde\Psi_0\,\rd\tau \to \frac{1}{\tau^2}\rd^{10}p\,\rd^{10}\Psi_0\,\rd^{10}\tilde\Psi_0\,\rd\tau\,.
\ee
Since the Pfaffians and $\delta$-functions are modular invariant, the only remaining factor comes from the ghost zero mode contribution $\rd z_1/(\rd z_0)^3$.  This produces the missing $\tau^2$ and renders the result modular invariant.


\section{Factorization}
\label{sec:factorize}

At genus one, there are two distinct factorization limits to consider when studying the IR behaviour of the NS-NS scattering amplitudes.  Heuristically, these correspond to the two ways in which the torus worldsheet can degenerate: either by pinching a cycle which reduces the torus to a Riemann sphere, or by pinching a cycle which factors the worldsheet into a sphere and another torus.  We refer to these as a \emph{non-separating} or \emph{separating} degeneration, respectively, and both can be understood as contributions from the boundary in the moduli space of curves $\overline{\cM}_{1,n}$ (c.f., \cite{Witten:2012bh} for a review).  

In the non-separating case, we approach a boundary divisor denoted by $\mathfrak{D}^{\mathrm{ns}}$, which looks like the moduli space of genus zero worldsheets with two additional punctures:
\begin{equation*}
 \mathfrak{D}^{\mathrm{ns}}\cong\overline{\cM}_{0,n+2}.
\end{equation*}
The separating degeneration corresponds to a divisor $\mathfrak{D}^{\mathrm{sep}}$ where the worldsheet pinches off a genus zero component $\Sigma_{L}\cong\CP^{1}$.  The $n$ marked points corresponding to the vertex operators distribute themselves between the two factors, with $n_L$ on $\Sigma_L$ and $n_R$ on $\Sigma_{R}$ such that $n_L+n_R=n$. This boundary divisor then looks like the product
\begin{equation*}
\mathfrak{D}^{\mathrm{sep}}\cong\overline{\cM}_{0,n_{L}+1}\times\overline{\cM}_{1,n_{R}+1}.
\end{equation*}

We confirm that in both factorization limits, the genus one amplitude develops a simple pole in the modulus transverse to the boundary divisor, as required by unitarity.  In addition, we also observe that in the non-separating degeneration, the amplitude is a rational function of the kinematic data, as appropriate for amplitudes in a field theory such as gravity.  This indicates that the various theta functions in the amplitude and partition function are actually subsumed by the sum over solutions to the scattering equations.  The situation in ordinary superstring theory is quite different, where factorized amplitudes are \emph{not} rational functions of kinematic data, and the Jacobi product expansion of theta functions builds an infinite series of modes on the string.

Of course, the unitary IR behavior of our formula (as well as the genus zero formulae of CHY) follows in a more abstract fashion simply by the properties of the worldsheet theory which produced it.  The worldsheet perspective on factorization allows us to deduce the IR behavior of amplitudes in this theory from basic geometric arguments, just as in ordinary string theory \cite{Polchinski:1988jq, Witten:2012bh} or twistor-string theory \cite{Adamo:2013tca}.  However, since our concern here is with the validity of the actual formula, it is important to derive the factorization behavior at the level of the amplitude itself.


\subsection{Pinching a non--separating cycle}
\label{sec:nonsep}

Pinching a non--separating cycle corresponds to approaching the non-separating boundary divisor $\mathfrak{D}^{\mathrm{ns}}\subset\overline{\cM}_{1,n}$, which is described by a degenerate limit of the complex structure $\tau$ for the torus worldsheet.  In particular, we need to consider the limit where $\mathrm{Im}\tau\rightarrow\infty$; to do this it is convenient to work with the alternative parameter $q=\e^{2\pi\im \tau}$ so that pinching the non--separating cycle is described by $q\rightarrow 0$.

As this cycle is pinched, it will be essential to understand how the various ingredients appearing in the expression for the amplitude behave. Using either their infinite sum or product representations, one can easily deduce that
\be\label{degb}
\eta(\tau)\sim q^{1/24}, \qquad \theta_{3}(0;\tau),\;\theta_{4}(0;\tau)\sim 1, \qquad \theta_{2}(0;\tau)\sim q^{1/8}, 
\ee
in the limit as $q\rightarrow 0$.  The behavior of the Szego kernel depends on the spin structure, and is apparent from \eqref{Szego} or can be rigorously derived using the sewing formalism for Riemann surfaces \cite{Yamada:1980, Tuite:2010mq}:
\be\label{degSzego}
S_{\bd \alpha}(z_{ij},\tau)\sim\left\{
\begin{array}{cc}
 \frac{\sqrt{\rd z_i}\;\sqrt{\rd z_j}}{z_i-z_j} & \mathrm{if}\;\;\bd\alpha=2 \\
 \kappa\times\sqrt{\rd z_i}\;\sqrt{\rd z_j} & \mathrm{otherwise}
\end{array}
\right. ,
\ee
where $\kappa$ is some constant.  On the right-hand side, we have abused notation by implicitly choosing an affine coordinate $z$ on the Riemann sphere; the appropriate coordinate system should always be evident from the context.  Similarly, we find that
\be\label{degSmod}
\tilde{S}_{1}(z_{i},z_{j};\tau)\sim \frac{\rd z_{i}}{z_{i}-z_{j}},
\ee
as $q\rightarrow 0$.

Upon pinching the non-separating cycle, the contribution to the amplitude from the odd spin structure vanishes since there are no odd spin structures on the sphere.  Hence, we only need to account for the behavior of $\cM_n^{1;\,{\rm even}}$ as $q\rightarrow 0$.  First, consider the behavior of $\mathrm{Pf}(M_{\bd\alpha})$, $\mathrm{Pf}(\widetilde{M}_{\bd\beta})$ in \eqref{evenamp}.  By \eqref{degSzego} and \eqref{degSmod}, it is clear that when $\bd\alpha=2$, the block entries of $M_{\bd\alpha}$ become:
\begin{equation*}
 A_{ij}=k_{i}\cdot k_{j}\frac{\sqrt{\rd z_i}\;\sqrt{\rd z_j}}{z_i-z_j}, \qquad B_{ij}=\epsilon_{i}\cdot\epsilon_{j}\frac{\sqrt{\rd z_i}\;\sqrt{\rd z_j}}{z_i-z_j}, \qquad C_{ij}=\epsilon_{i}\cdot k_{j}\frac{\sqrt{\rd z_i}\;\sqrt{\rd z_j}}{z_i-z_j},
\end{equation*}
which are the expected entries at genus zero \cite{Cachazo:2013hca, Mason:2013sva}.  The only subtlety is in the diagonal entries of the $C$-block, which read:
\begin{equation*}
 C_{ii}|_{q\rightarrow0}=-\sum_{j\neq i}\frac{\epsilon_{i}\cdot k_{j}}{z_i-z_j}\rd z_i\;+\epsilon_{i}\cdot p|_{q\rightarrow 0}\;\rd z_{i},
\end{equation*}
where $p_{\mu}\rd z_{i}$ is the zero mode of $P_{\mu}(z_i)$ on the torus.  On the boundary divisor $\mathfrak{D}^{\mathrm{ns}}$, a global holomorphic differential (such as $p_{\mu}\rd z_{i}$) degenerates into a meromorphic differential on the sphere with simple poles at the two new marked points, having equal and opposite residues at those points (c.f., \cite{Faybook}).  Calling this residue $k_{\mu}$, and denoting the two new marked points as $z_{a},z_{b}\in\CP^{1}$, we find:
\begin{equation*}
C_{ii}|_{q\rightarrow0}=\left(-\sum_{j\neq i}\frac{\epsilon_{i}\cdot k_{j}}{z_i-z_j}+\frac{\epsilon_{i}\cdot k}{z_{i}-z_{a}}-\frac{\epsilon_{i}\cdot k}{z_{i}-z_{b}}\right)\rd z_{i}=C_{ii}^{n+2},
\end{equation*}
which is the diagonal entry for the $C$-block with $n+2$ particles, two of which have equal and opposite momentum.  The story for $\widetilde{M}_{\bd\alpha}$ is identical.

Hence, we see that
\be\label{ns1}
\mathrm{Pf}(M_{2}),\;\mathrm{Pf}(\widetilde{M}_{2})\xrightarrow{q\rightarrow 0} \mathrm{Pf}(M^{ab}_{ab}),\;\mathrm{Pf}(\widetilde{M}^{ab}_{ab}),
\ee
where $M^{ab}_{ab}$ is the matrix whose entries are the same as in the genus zero case for $n+2$ particles, with rows and columns corresponding to the new external states at $z_{a},z_{b}$ (and with momentum $k_{\mu}$, $-k_{\mu}$) removed.  Note that unlike boson scattering amplitudes at genus zero, the rank of the Pfaffian is un-changed.  For the other two even spin structures, the matrices $M_{\bd\alpha}$, $\widetilde{M}_{\bd\alpha}$ do not approach recognizable structures.  However, we will see that these contributions actually cancel due to the GSO projection.

At this point, we note that the only factors in $\cM_n^{1;\,{\rm even}}$ which encode the spin structure and potential $q$-dependence are
\begin{multline}\label{ns2}
 \rd \tau\;\sum_{\bd\alpha;\bd\beta} (-1)^{\bd\alpha+\bd\beta} Z_{\bd\alpha;\bd\beta}(\tau){\rm Pf}(M_{\bd\alpha})\,{\rm Pf}(\widetilde{M}_{\bd\beta}) \\
=\frac{1}{2\pi i}\frac{\rd q}{q}\sum_{\bd\alpha;\bd\beta} (-1)^{\bd\alpha+\bd\beta} \frac{\theta_{\bd\alpha}(0;\tau)^{4}\;\theta_{\bd\beta}(0;\tau)^4}{\eta(\tau)^{24}} {\rm Pf}(M_{\bd\alpha})\,{\rm Pf}(\widetilde{M}_{\bd\beta}).
\end{multline}
Using the leading behavior given by \eqref{degb}, we see that as $q\rightarrow 0$ this sum looks like
\be\label{ns3}
\frac{\rd q}{q^{2}}\sum_{\bd\beta} (-1)^{\bd\beta} \theta_{\bd\beta}(0;\tau)^4\;{\rm Pf}(\widetilde{M}_{\bd\beta})\left[q^{1/2}{\rm Pf}(M_{2})-{\rm Pf}(M_{3})+{\rm Pf}(M_{4})\right], 
\ee
which appears to have a tachyonic double pole in $q$.  But as $q\rightarrow 0$, we know that ${\rm Pf}(M_{3})={\rm Pf}(M_{4})$, so the last two terms in \eqref{ns3} cancel with each other via the GSO projection.  

The same argument works for the sum over $\bd\beta$, leading to power of $q$ in the numerator from the only surviving term where $\bd\alpha=\bd\beta=2$.  Hence, close to the boundary divisor $\mathfrak{D}^{\mathrm{ns}}$ the contribution to the measure from \eqref{ns2} is given by:
\be\label{ns4}
\rd \tau\;\sum_{\bd\alpha;\bd\beta} (-1)^{\bd\alpha+\bd\beta} Z_{\bd\alpha;\bd\beta}(\tau){\rm Pf}(M_{\bd\alpha})\,{\rm Pf}(\widetilde{M}_{\bd\beta})\sim \frac{\rd q}{q} \mathrm{Pf}(M^{ab}_{ab})\;\mathrm{Pf}(\widetilde{M}^{ab}_{ab}).
\ee
This is in direct analogy with the role of the GSO projection in ordinary string theory: a generic term in $\cM_{n}^{1;\,{\rm even}}$ has a tachyonic double pole in the modulus $q$ as we pinch the non-separating cycle, but the sum over spin structures (with appropriate signs dictated by modular invariance) cancels these double poles and leaves only the simple pole consistent with unitarity.

The last piece of the amplitude we need to analyse in this factorization limit are the scattering equations enforced by
\begin{align}\label{scaeq}
\bar{\delta}(P(z_1)^2)\prod_{j=2}^{n}\bar{\delta}(k_j\cdot P(z_j)).
\end{align}
The role of these equations is to set to zero the meromorphic quadratic differential $P^2(z)$ by imposing that any possible
pole has zero residue \eqref{vertexseq} and that its value at a point is zero \eqref{moduliseq}.  As we approach the boundary divisor, the equation for the residues of $P^{2}(z)$ reduces to the familiar form of the tree-level scattering equations
\begin{align}\label{scaeqfac}
 k_i\cdot P(z_i)= \frac{k_i\cdot k}{z_i-z_a}-\frac{k_i\cdot k}{z_i-z_b}+\sum_{j\neq i}\frac{k_i\cdot k_j}{z_i-z_j},
\end{align}
where two new particles were created at points $z_a,z_b$ with equal and opposite momentum $k$. Taking this factorization
limit leaves us with an $(n+2)$-point tree amplitude which should come with $n-1$ scattering equations, which is precisely the number of equations given for each choice of $i$ in \eqref{scaeqfac}.  As usual in the factorization limit we insert operators $c\tilde c$ which
create punctures so the states inserted at these points are fixed; hence we don't get scattering equations for the particles
inserted at $z_a,z_b$.

On the support of \eqref{scaeqfac} the remaining scattering equation becomes
\begin{align}\label{on_shellscaeq}
P^2(z_1)=p^2\rd z_1^2=k^2\rd z_1^2\left(\frac{z_a-z_b}{(z_1-z_a)(z_1-z_b)}\right)^2=0 \, ,
\end{align}
which forces the momentum running through the cut to be on-shell, with $\{z_1,z_a,z_b\}$ fixed by the $\SL(2,\mathbb{C})$ freedom on the degenerate worldsheet. 

\medskip

We take this opportunity to note that for generic values of the modular parameter $\tau$, $\bar\delta(P^2)$ does \emph{not} constrain $p_{\mu}$ to be null.  If this were true, then the loop momentum would always be constrained to be on-shell.  For a generic value of $\tau$, we can use the remaining $n-1$ scattering equations and momentum conservation to write \eqref{moduliseq} as
\begin{align}
 P^2(z_1)=p^2\;\rd z^{2}+\sum_{j\neq i}k_j\cdot k_i\; f(z_i,z_j,\tau)\;\rd z^{2}\, ,
\end{align}
where the function $f(z_i,z_j,\tau)$ is smooth and has no singularity when $x_i\rightarrow x_j$.  Furthermore, when $\mathrm{Im}\tau\rightarrow\infty$, $f$ approaches a constant independent of the worldsheet coordinates.  By momentum conservation, this means that $P^{2}(z_1)\rightarrow p^{2}$ as we pinch the non-separating cycle.  Hence, the degeneration parameter $q$ is directly related to the off-shellness of the internal loop momentum.

This implies that in general the scattering equation \eqref{moduliseq} can be seen as fixing the integration over $\tau$, leaving a loop integral over the non-compact space of $P$ zero modes. Integrating over this space might introduce divergences which are absent from
string theory amplitudes but are expected from a theory which gives field theory amplitudes.  We can also interpret this equation as reducing the integral over the $P$ zero modes to some hypersurface parametrized by $\tau$. The moduli of the Riemann surface then can be seen as an off-shellness parameter for the loop momentum and we retain the interpretation that the target space is ambitwistor space.

\medskip

We have seen that when a non-separating cycle is pinched a pole of order one appears and the amplitude factorizes in terms
of an expression on a genus zero worldsheet with two additional particles of equal and opposite null momenta. This is integrated over the
phase space of the on-shell loop momenta and summed over all possible intermediate states.  Critically, the integrand of the result is a rational function of kinematic invariants, as expected for field theory amplitudes; the various elliptic functions only contribute to the simple pole rather than adding higher mode dependence as in ordinary superstring theory.  This is true for the same reason that the integrand of the tree-level expression for graviton amplitudes is a rational function. 

In this factorized amplitude, the intermediate states could be gravitons or gravitinos. While there is a compact expression for $n$-graviton scattering that could be used to check the above formula, we lack a similarly simple expression for 2-gravitino and $(n-2)$-graviton scattering. Nevertheless the result of this factorization limit seems to imply that a simple expression for such amplitudes exists. Perhaps a formalism
which makes target-space supersymmetry manifest as in \cite{Berkovits:2013xba} could be used to find such expressions.

\subsection{Pinching a separating cycle}
\label{sec:sep}

Pinching a separating cycle on the genus one worldsheet factors off a Riemann sphere $\Sigma_{L}\cong\CP^{1}$ as we approach the boundary divisor $\mathfrak{D}^{\mathrm{sep}}$.  In this case, the degeneration of the worldsheet has nothing to do with the modular parameter $\tau$; instead, it corresponds to a set of $n_{L}$ of the external vertex operators becoming very close to each other.  A conformally equivalent situation is that these $n_{L}$ insertions are on a sphere $\Sigma_L$ which is connected to the torus $\Sigma_R$ by a long tube.

In the neighborhood of this tube, we can model the worldsheet by
\be\label{sep1}
(z_{L}-w)(z_{R}-y)=s,
\ee
where $z_{L}$ is a local coordinate on $\Sigma_L$ and $z_R$ is a local coordinate on $\Sigma_R$.\footnote{Once again, we will leave the choice of a coordinate system on $\Sigma_L$ or $\Sigma_R$ implicit from now on.}  Clearly, $s$ acts as a modulus for the length of the tube connecting the two branches, and as $s\rightarrow 0$ the worldsheet separates into $\Sigma_{L}\cup\Sigma_R$, joined at the points $z_L=w$ and $z_R=y$ (see \cite{Witten:2012bh} for a review).  Thus, we can think of $s$ as a modulus transverse to the boundary divisor $\mathfrak{D}^{\mathrm{sep}}\subset\overline{\cM}_{1,n}$.  We are interested in the behavior of the genus one scattering amplitude as we approach this boundary.

Unfortunately, the expression for the $g=1$ amplitude computed in \ref{sec:1loopamps} is not optimal for studying the separating degeneration.  This is because we calculated the amplitude in a picture with no insertions of $\delta(\gamma)$ or $\delta(\tilde{\gamma})$; this was natural because there are no zero modes of the superconformal ghosts which need to be fixed at genus one.  However, upon pinching the separating cycle we produce the branch $\Sigma_L$ on which $\gamma$ and $\tilde{\gamma}$ have two zero modes each.  In other words, the two worldsheets produced by the separating degeneration have different numbers of fermionic moduli.  The new states we expect to appear at $w\in\Sigma_L$ and $y\in\Sigma_R$ should be represented by fixed vertex operators (i.e., with picture number $-1$), which is un-natural from the perspective of the picture used in section \ref{sec:1loopamps}.  In other words, the use of integrated vertex operators corresponds to a choice of gauge which makes pinching a separating cycle difficult.

This issue is familiar from the conventional RNS superstring: at arbitrary genus, amplitudes are easiest to compute using a mixture of fixed and integrated vertex operators appropriate to the number of zero modes in the superconformal ghost system.  At the level of the integrand (i.e., before performing the moduli integrals), this expression is optimal in the sense that it minimizes the number of picture changing insertions and behaves appropriately under all non-separating factorizations and all separating factorizations for which the resulting worldsheets have the same number of fermionic zero modes.\footnote{For example, at genus two the expression will factorize correctly for a non-separating degeneration as well as the separating degeneration that results in two tori (c.f., \cite{DHoker:2002gw}).}  However, this choice of picture is un-natural for generic worldsheet degenerations where new states will appear in the fixed picture, making it difficult to isolate the IR behavior of the amplitude.  

The solution to this issue is to represent all external states by fixed vertex operators at the expense of introducing an appropriate number of picture changing operators.  The resulting amplitude--while appearing superficially different from an expression obtained with integrated vertex operators--will be independent of the PCO insertions and in fact \emph{equal} to the alternative expression (although proving this in specific examples can be difficult).  The amplitude in this all-fixed picture is naturally suited to studying all boundary divisors in the moduli space since all external states are on the same footing as new states which appear in the factorization channel.  Another way of seeing this is by considering the worldsheet perspective on factorization, where it is essential to work in the all-fixed picture (see \cite{Witten:2012bh, Adamo:2013tca} for more details).

At genus one in even spin structure, this means that we should compute the NS--NS sector scattering amplitude from the worldsheet correlation function:
\be\label{sep2}
\cM^{1;\;\rm{even}}_{n}=\left\la \prod_{i=1}^{n}c_{i}\tilde{c}_{i}U_{i} \prod_{a=1}^{n}\Upsilon_{a}\widetilde{\Upsilon}_{a}\;\prod_{r=1}^{n-1}(b_{r}|\mu_{r})\;(\tilde{b}_{r}|\mu_{r})\bar{\delta}\left(\int_{\Sigma}\mu_{r}\;P^{2}\right)\right\ra,
\ee  
where we use the short-hand
\begin{equation*}
(b_{r}|\mu_{r})=\int_{\Sigma}b_{r}\wedge\mu_{r}\, ,
\end{equation*}
for the measure on the moduli space.  

The resulting amplitude can be computed in much the same way as our previous expression.  In an even spin structure, we find:
\begin{multline}\label{sepeven}
\cM^{1;\;\mathrm{even}}_{n}=\delta^{10}\left(\sum_i k_i\right)\int \rd^{10}p\wedge\rd\tau \wedge \bar\delta\left(P^2(z_1)\right)\,\prod_{i=2}^n \bar\delta(k_i\cdot P(z_i)) \\
	\times\sum_{\bd\alpha;\bd\beta} (-1)^{\bd\alpha+\bd\beta} Z_{\bd\alpha;\bd\beta}(\tau) \frac{{\rm Pf}(\mathsf{M}_{\bd\alpha})}{|\mathsf{R}_{\bd\alpha}|}\frac{{\rm Pf}(\widetilde{\mathsf{M}}_{\bd\beta})}{|\widetilde{\mathsf{R}}_{\bd\beta}|},
\end{multline}
where the partition function $Z_{\bd\alpha;\bd\beta}(\tau)$ is as in \eqref{partition}.  The skew-symmetric $2n\times2n$ matrix $\mathsf{M}_{\bd\alpha}$ arises from the matter systems, is analogous to the matrix $M_{\bd\alpha}$ appearing in \eqref{evenamp}, and has a block decomposition
\begin{equation*}
\mathsf{M}_{\bd\alpha}=\left(
\begin{array}{cc}
 \mathsf{A} & -\mathsf{C}^{\mathrm{T}} \\
\mathsf{C} & \mathsf{B}
\end{array}\right).
\end{equation*}
Entries of the $\mathsf{A}$-block are indexed by the locations of the PCOs, which we denote as $x_{a},x_{b}\in\Sigma$, for $a,b=1,\ldots,n$:
\begin{multline}\label{sepA}
\mathsf{A}_{ab}=S_{\bd\alpha}(x_{ab};\tau)\left(\sum_{i,j=1}^{n}k_{i}\cdot k_{j}\;\tilde{S}_{1}(x_{a},z_i;\tau)\;\tilde{S}_{1}(x_{b},z_j;\tau) +\sum_{i=1}^{n}k_{i}\cdot p\; \rd x_{b}\;\tilde{S}_{1}(x_{a},z_{i};\tau) \right. \\
\left. +\sum_{j=1}^{n}p\cdot k_{j}\;\rd x_{a}\;\tilde{S}_{1}(x_{b},z_{j};\tau)+p^{2}\;\rd x_{a}\;\rd x_{b}\right),
\end{multline}
with $\mathsf{A}_{aa}=0$.  The entries of the $\mathsf{B}$-block are indexed by the vertex operator locations, and are identical to those in \eqref{ABCg=1}:
\be\label{sepB}
\mathsf{B}_{ij}=\epsilon_{i}\cdot\epsilon_{j}\;S_{\bd\alpha}(z_{ij};\tau), \qquad \mathsf{B}_{ii}=0.
\ee
Finally, the rows of the $\mathsf{C}$-block are indexed by the vertex operators, while its columns are indexed by the PCOs:
\be\label{sepC}
\mathsf{C}_{ia}=S_{\bd\alpha}(x_{a}-z_{i};\tau)\left(\sum_{j=1}^{n}\epsilon_{i}\cdot k_{j}\;\tilde{S}_{1}(x_{a},z_{j};\tau)+\epsilon_{i}\cdot p\;\rd x_{a}\right).
\ee

A determinant of the $n\times n$ matrix $\mathrm{R}_{\bd\alpha}$ arises in the denominator due to the correlator in the $\beta\gamma$-system.  This is the expected bosonic `Slater determinant' (c.f., \cite{Witten:2012bh}, Section 10) whose entries are composed of the propagators between the $\gamma$ insertions for vertex operators and the $\beta$ insertions for the PCOs:
\be\label{sepR}
\mathsf{R}_{ia}=S_{\bd\alpha}(z_{i}-x_{a};\tau)\frac{\rd x_{a}}{\rd z_{i}}.
\ee
Of course, then entries of $\widetilde{\mathsf{M}}_{\bd\beta}$ and $\widetilde{\mathsf{R}}_{\bd\beta}$ are exactly the same, except for the spin structure and polarization vectors.

At first, it may appear that \eqref{sepeven} cannot be equivalent to our earlier expression \eqref{evenamp}: not only are the various Pfaffians different, but there are also novel Slater determinants as well as apparent dependence on the locations of the PCOs.  Of course, this answer must be independent of the locations $x_{a}$, but there appear to be various poles in $\mathsf{M}_{\bd\alpha}$ and $\widetilde{\mathsf{M}}_{\bd\beta}$ as these locations coincide with the external operator locations $z_i$.  However, by carefully considering the limit where $x_{i}\rightarrow z_{i}$, it can be shown that all these apparent singularities vanish, and the resulting expression is in fact \emph{equal to} \eqref{evenamp}.  By Liouville's theorem, this means that \eqref{sepeven} and \eqref{evenamp} are equivalent representations of the even spin structure contribution to the amplitude!  A similar story exists for the odd spin structure, although we will not present it explicitly here.

So \eqref{sepeven} provides us with an expression for the genus one amplitude in which all external states are on the same footing.  This allows us to pinch the separating cycle using the local model \eqref{sep1}.  All the ingredients in the amplitude which are associated uniquely with the torus simply remain on the $\Sigma_R$ factor without contributing any dependence on the parameter $s$.  In particular, the integrals over $\rd^{10}p$ and $\rd\tau$, as well as $Z_{\bd\alpha;\bd\beta}$ simply move onto $\Sigma_R$ as $s\rightarrow 0$.  The odd spin structure also contributes nothing to the $\Sigma_L$ branch since there is no odd spin structure on the sphere.

As we pinch the separating cycle, we let $n_L$ of the vertex operators move onto $\Sigma_L$, while the remaining $n_R=n-n_L$ remain on $\Sigma_R$.  The PCO locations also divide themselves between the two factors; in order for the result to be non-vanishing, we must have $n_L-1$ of the $x_a$ on $\Sigma_L$ and $n_R+1$ on $\Sigma_R$.  Near the boundary divisor, there is a natural identification of three of the moduli in play: the modulus $s$, and the locations of the two new fixed points $w,y$.  These will contribute to the overall measure as \cite{Witten:2012bh}
\be\label{sep3}
\rd w\;\rd y\;\frac{\rd s}{s^2},
\ee
by the scaling properties of \eqref{sep1}.  We expect that the form degrees in $w,y$ will be absorbed by the various Pfaffians and scattering equations, so we begin with an insertion of $s^{-2}\rd s$ as we approach the boundary divisor.

As the worldsheet degenerates, the scattering equations likewise become degenerate.  Since this degeneration is practically identical to the situation for factorization at genus zero \cite{Cachazo:2013gna}, we will be rather brief here.  Recall that zero-modes of $P_{\mu}$ solve the equation:
\begin{equation*}
 \dbar P_{\mu}(z)=2\pi\im\,\rd z\wedge\rd\bar{z}\;\sum_{i=1}^{n}k_{i\;\mu}\;\delta^{2}(z-z_i).
\end{equation*}
On a genus zero curve, this equation has no homogeneous solution, while on the torus it has the homogeneous solution $p_{\mu}\rd z$.  As we approach the separating divisor, $P_{\mu}$ develops a new homogeneous term on each factor $\Sigma_L$, $\Sigma_R$ which must have a simple pole at the new marked point, with opposite residue on the two factors (c.f., \cite{Faybook}).  This residue is then integrated as part of the integral over zero-modes, and is interpreted as the momentum flowing through the cut.  In particular, this means that we have:
\begin{eqnarray}
P_{\mu}(z)|_{\Sigma_L}\rightarrow -\frac{k_{R\;\mu}}{z-w}\rd z+\sum_{i\in L}\frac{k_{i\;\mu}}{z-z_i}\rd z, \label{sep4} \\
P_{\mu}(z)|_{\Sigma_R}\rightarrow  p_{\mu}\rd z+k_{R\;\mu}\tilde{S}_{1}(z,y;\tau) +\sum_{j\in R}k_{j\;\mu}\tilde{S}_{1}(z,z_j;\tau) \label{sep5}.
\end{eqnarray}

Now, the original set of $n-1$ scattering equations splits into a set of $n_{L}-2$ scattering equations on $\Sigma_L$ and $n_{R}$ scattering equations on $\Sigma_{R}$.  The remaining scattering equation degenerates into a delta function enforcing momentum conservation on each factor as $s\rightarrow 0$. Without loss of generality, we can assume that $z_1,z_2\in\Sigma_L$ which leaves us with:
\begin{multline}\label{sep6}
\bar\delta\left(P^2(z_1)\right)\prod_{i=2}^n \bar\delta(k_i\cdot P(z_i))\longrightarrow \frac{\rd w\;\rd y}{s} \; \bar{\delta}\left(s\mathcal{F}+k^{2}_{R}\right)\\
 \times\frac{z_{12}z_{2w}z_{w1}}{\rd z_{1}\;\rd z_{2}\rd w}\prod_{i\in L\setminus\{1,2\}}\bar{\delta}\left(k_{i}\cdot P(z_i)\right)\;\bar{\delta}\left(P^2(y)\right) \prod_{j\in R}\bar{\delta}\left(k_{j}\cdot P(z_j)\right), 
\end{multline}
where $\mathcal{F}$ is some rational function of the $k_i$ and $z_i$ \cite{Cachazo:2013gna}.  The first factor on the right-hand side ensures that the resulting expression has the correct homogeneity and form degrees required by \eqref{sep1}.  In conjunction with \eqref{sep4}--\eqref{sep5}, we find the scattering equations for $n_L+1$ external particles on $\Sigma_L$ and for $n_R+1$ external particles on $\Sigma_R$, as desired.

Now let us turn to the behavior of the Pfaffians as $s\rightarrow 0$.  Every entry in $\mathsf{M}_{\bd\alpha}$ falls into one of two classes: either both of its indices are on the same side of the separating cycle, or they are on different sides.  If $z,z'\in\Sigma_L$, then as $s\rightarrow0$ the Szego kernel $S_{\bd\alpha}(z-z';\tau)$ simply reduces to the Szego kernel on $\Sigma_L$, and similarly for $z,z'\in\Sigma_R$ \cite{Yamada:1980, Tuite:2010mq}.  

On the other hand, when $z\in\Sigma_L$ and $z'\in\Sigma_R$, homogeneity and conformal invariance dictate that the Szego kernel behaves like
\be\label{sepSz1}
S_{\bd\alpha}(z-z';\tau)=\frac{\sqrt{s}}{\sqrt{\rd w}\sqrt{\rd y}}\frac{\sqrt{\rd z}\sqrt{\rd w}}{z-w}\;S_{\bd\alpha}(y-z';\tau) +O(s^{3/2}),
\ee
as $s\rightarrow 0$.  Similar reasoning dictates that the propagator $\tilde{S}_1$ behaves as
\be\label{sepSz2}
\tilde{S}_{1}(z,z';\tau)=\frac{s}{\rd y} \frac{\rd z}{z-w}\;\tilde{S}_{1}(y, z';\tau)+O(s^2),
\ee
in this situation.  

This allows us to determine the behavior of the entries in $\mathsf{M}_{\bd\alpha}$ in the $s\rightarrow 0$ limit.  For instance, if $x_{a},x_{b}\in\Sigma_L$ then
\be\label{sepAL}
\mathsf{A}_{ab}= \frac{\sqrt{\rd x_{a}}\sqrt{\rd x_{b}}}{x_{a}-x_{b}}\sum_{i,j\in L\cup\{w\}}k_{i}\cdot k_{j}\frac{\rd x_{a}\;\rd x_{b}}{(x_{a}-z_i)(x_b-z_j)}+O(s).
\ee
Using \eqref{sep4}--\eqref{sep5} in conjunction with \eqref{sepSz1}--\eqref{sepSz2} it is easy to see that for a general entry in $\mathsf{M}_{\bd\alpha}$, we have
\be\label{sep7}
(\mathsf{M}_{\bd\alpha})_{i_L j_L}\rightarrow (\mathsf{M}^{L})_{i_L j_L}, \qquad (\mathsf{M}_{\bd\alpha})_{i_R j_R}\rightarrow (\mathsf{M}^{R}_{\bd\alpha})_{i_R j_R},
\ee
where $\mathsf{M}^{L}$ is the matrix for the genus zero amplitude on $\Sigma_L$ with external particles in $L\cup\{a\}$ and $\mathsf{M}^{R}_{\bd\alpha}$ is the matrix for the genus one amplitude on $\Sigma_R$ with external particles in $R\cup\{b\}$.    

But what about the entries of $\mathsf{M}_{\bd\alpha}$ which tie together locations on opposite sides of the separating cycle?  A simple calculation reveals that for $x_a\in\Sigma_L$, $x_{b}\in\Sigma_R$,
\begin{multline}\label{sepALR}
\mathsf{A}_{ab}= \frac{\sqrt{s}}{\sqrt{\rd w}\sqrt{\rd y}}\frac{\sqrt{\rd x_{a}}\sqrt{\rd w}}{x_a-w}S_{\bd\alpha}(y-x_b;\tau) \\
\times\left(\sum_{i\in L\cup\{w\}}\sum_{j\in R\cup\{y\}}k_{i}\cdot k_{j} \frac{\rd x_a}{x_a-z_i}\tilde{S}_{1}(x_b,z_j;\tau) +\sum_{i\in L\cup\{w\}}k_{i}\cdot p\;\rd x_{b}\frac{\rd x_a}{x_a-z_i}\right) +O(s^{3/2})
\end{multline}
as $s\rightarrow 0$.  Likewise, for $x_a\in\Sigma_L$ and $z_i\in\Sigma_R$ we find
\be\label{sepCRL}
\mathsf{C}_{ia}=\frac{\sqrt{s}}{\sqrt{\rd w}\sqrt{\rd y}}\frac{\sqrt{\rd x_{a}}\sqrt{\rd w}}{x_a-w}S_{\bd\alpha}(y-z_i;\tau)\sum_{j\in L\cup\{w\}}\epsilon_{i}\cdot k_{j}\frac{\rd x_a}{x_a-z_j}+O(s^{3/2}),
\ee
and for $z_{i}\in\Sigma_L$, $z_j\in\Sigma_R$,
\be\label{sepBRL}
\mathsf{B}_{ij}=\frac{\sqrt{s}}{\sqrt{\rd w}\sqrt{\rd y}}\frac{\sqrt{\rd z_{i}}\sqrt{\rd w}}{z_i-w}S_{\bd\alpha}(y-z_i;\tau)\;\epsilon_{i}\cdot\epsilon_{j}+O(s^{3/2}).
\ee

In each of these entries, we have a product $e_i\cdot e_j$, where $e^{\mu}$ is either a momentum or polarization vector.  The completeness relation allows us to write these contractions in terms of polarization vectors:
\begin{equation*}
e_{i}\cdot e_{j}=e_{i}^{\mu}\;e_{j}^{\nu}\left(\sum_{\epsilon_I}\epsilon_{a\;\mu}\epsilon_{b\;\nu}-\frac{k_{R\;\mu}k_{R\;\nu}}{k_{R}^{2}}\right), 
\end{equation*}
where the sum runs over the possible polarizations of the internal particle.  The second term in this expression is actually just a gauge transformation so it can be neglected.  Upon inspecting \eqref{sepALR}--\eqref{sepBRL}, we can see that the completeness relation actually generates all the entries in the $(2w)^{\rm th}$ row and column of $\mathsf{M}^{L}$ as well as the $(2y)^{\rm th}$ row and column of $\mathsf{M}^{R}_{\bd\alpha}$, up to an overall factor proportional to $\sqrt{s}$.

Using the basic properties of Pfaffians, we now deduce the factorization behavior of $\mathrm{Pf}(\mathsf{M}_{\bd\alpha})$ as the separating cycle is pinched:
\be\label{sepPf}
{\rm Pf}(\mathsf{M}_{\bd\alpha})\rightarrow \frac{\sqrt{s}}{\sqrt{\rd w}\sqrt{\rd y}} \mathrm{Pf}(\mathsf{M}^{L})\;\mathrm{Pf}(\mathsf{M}^{R}_{\bd\alpha}),
\ee
where $\mathsf{M}^{L}$ is the $2n_{L}\times 2n_{L}$ matrix at genus zero and $\mathsf{M}^{R}_{\bd\alpha}$ is the $2(n_R+1)\times 2(n_R+1)$ matrix at genus one.  The final ingredient is given by the factorization of the determinant $|\mathsf{R}_{\bd\alpha}|$, which is guaranteed by the properties of the $\beta\gamma$-system.\footnote{This behavior is universal for the superconformal ghost system, or for any general Slater determinant, in ordinary string theory as well as the ambitwistor string.}  In particular, we have:
\be\label{sepRL}
|\mathsf{R}_{\bd\alpha}|\rightarrow \frac{1}{\sqrt{s}}|\mathsf{R}^{L}|\;|\mathsf{R}^{R}_{\bd\alpha}|,
\ee
for the appropriate $(n_{L}+1)\times (n_L+1)$ Slater determinant on $\Sigma_L$ and $(n_R+1)\times(n_R+1)$ determinant on $\Sigma_R$.  The factor of $s^{-1/2}$ ensures the appropriate homogeneity, since there is now a row corresponding to $w$ in $\mathsf{R}^{L}$ and a row corresponding to $y$ in $\mathsf{R}^{R}_{\bd\alpha}$.

Pulling all the pieces together, we find that near the separating boundary divisor the genus one amplitude looks like:
\begin{multline}
 \int \frac{z_{12}z_{2w}z_{w1}}{\rd z_{1}\;\rd z_{2}\rd w}\prod_{i\in L\setminus\{1,2\}}\bar{\delta}\left(k_{i}\cdot P(z_i)\right) \frac{\mathrm{Pf}(\mathsf{M}^{L})}{|\mathsf{R}^{L}|} \frac{{\rm Pf}(\widetilde{\mathsf{M}}^{L})}{|\widetilde{\mathsf{R}}^{L}|} \: \frac{\rd s}{s}\;\bar{\delta}\left(s\mathcal{F}+k^{2}_{R}\right) \\
\rd^{10}p\;\rd \tau\;\bar{\delta}\left(P^2(y)\right)\prod_{j\in R}\bar{\delta}\left(k_{j}\cdot P(z_j)\right)\sum_{\bd\alpha;\bd\beta}(-1)^{\bd\alpha+\bd\beta}Z_{\bd\alpha;\bd\beta}(\tau)\;\frac{\mathrm{Pf}(\mathsf{M}^{R}_{\bd\alpha})}{|\mathsf{R}^{R}_{\bd\alpha}|} \frac{\mathrm{Pf}(\widetilde{\mathsf{M}}^{R}_{\bd\beta})}{|\widetilde{\mathsf{R}}^{R}_{\bd\beta}|}.
\end{multline}
As expected, there is only a simple pole in the degeneration modulus $s$; taking the residue of this pole sets the momentum flowing across the cut to be null ($k^2_{R}=0$), and it is easy to show that the resulting on-shell amplitudes for $\Sigma_L$ and $\Sigma_R$ are equivalent to the genus zero NS--NS formula and \eqref{sepeven} respectively.  

Hence, the genus one amplitude of the ambitwistor string factorizes correctly in the separating channel.  Note that in this case the resulting amplitudes were identified as the tree-level and one-loop all-boson amplitudes.  This is because the Ramond sector cannot contribute to the separating degeneration, since the resulting amplitudes would have only one external fermion and therefore vanish.


\section{Conclusions}
\label{sec:conclusions}

In this paper we have used the ambitwistor string of~\cite{Mason:2013sva} to extend the formul\ae\ for $n$-particle gravitational scattering amplitudes found in~\cite{Cachazo:2013hca} in several directions. Firstly, we showed that, in close analogy to the usual string, the ambitwistor string contains Ramond--Neveu Schwarz vertex operators describing fermionic states in the target space. As usual, these can be the gravitinos of the type II supergravity, or gravitinos and gauginos in a heterotic model. We studied the simplest scattering amplitudes involving these fermionic states, showing explicitly that they agree with the expected tree level supergravity amplitudes. The type II models also contain massless $p$-form fields in the Ramond--Ramond sector, and as usual these have $p$ even or odd depending on the choice of GSO projections. However, due to triviality of the $XX$ OPE, the ambitwistor string contains no $\alpha'$ excitations, so the complete spectrum of the type IIA or type IIB ambitwistor string is just type II supergravity in ten dimensions. 

We then investigated the genus one correction to these scattering amplitudes. The integral over the moduli space of $n$-pointed elliptic curves was again shown to be completely fixed by a genus one version of the scattering equations. These equations involved $n\!-\!1$ constraints on the residues of $P^2$ at the vertex operators, and also a further constraint on $P^2$ itself that originated from the measure of integration over the moduli of the worldsheet gauge field $e$ responsible for quotienting the target space from $T^*M$ to ambitwistor space. We also showed the loop integral $\rd^{10}p$ itself emerges naturally in our formalism as the integral over zero modes of the $P_\mu$ field. This is somewhat similar to the origin of the loop integral in the chiral factorization theorems of D'Hoker and Phong~\cite{DHoker:1989ai}, although here the zero mode $p_\mu$ is an independent field, unrelated to periods of $\del X^\mu$. In particular the definition of these zero modes does not appear to require a choice of homology cycles; we merely pick any basis of $H^0(\Sigma,K_\Sigma)$.

We computed the ambitwistor string partition function and showed that it is modular invariant --- a highly non--trivial result for a chiral theory. In the case of type II ambitwistor strings, the partition function vanishes as a consequence of target space supersymmetry (as usual), but a non--supersymmetric type 0 theory also appears to exist. It would be interesting to investigate this theory further, particularly because unlike the usual string, it does not appear to contain a tachyon in its (strictly massless) spectrum. 

We then computed the $n$-point $g=1$ correlators for states in the NS--NS sector of supergravity, showing that just as at tree level, these correlation functions may be represented in terms of Pfaffians. We checked that these formul\ae\ have the expected behaviour under factorization of the worldsheet, both in separating and non--separating channels.

\medskip

This paper leaves open many unanswered questions. Firstly, since the integral over $\cM_{1,n}$ is here interpreted as providing the one-loop {\it integrand} of supergravity, with the loop integral left to be done, we would expect that the worldsheet correlation function becomes simply a {\it rational} function of the external and loop momenta. In particular, the presence of elliptic functions, while completely natural from the point of view of correlation functions on a torus, is completely the wrong category for what we expect in pure supergravity. For instance, the Jacobi product expansion of theta functions is usually interpreted as describing the contribution of all the higher string modes to the correlation function.  The only ray of hope seems to be that the $g=1$ scattering equations do not directly fix the worldsheet coordinates in terms of the external and loop momenta, but rather fix elliptic functions of the worldsheet coordinates in terms of these momenta. Thus, we appear to require that a miraculous simplification should occur, in which all trace of elliptic functions disappears, after summing the correlator over all solutions of the $g=1$ scattering equations (with the appropriate Jacobian). 

It is not at all clear to us how this actually transpires, and for example whether it occurs only after also summing over spin structures. However, the non-separating factorization channel (corresponding to the single cut of the loop amplitude) \emph{does} lead to a rational function of the kinematic data, which is strong evidence in favor of such a simplification. This indicates that (unlike conventional superstring theory) no trace of the elliptic functions remains in the factorization limit, and the scattering equations hopefully perform this simplification in the interior of the moduli space as well.  It would be fascinating to see this explicitly, even in the case of the $n=4$ particle amplitude. We remark that a similar simplification must also be at work in the $\cN=8$ twistor string of~\cite{Skinner:2013xp}. 

Going further, it would be important to actually {\it solve} the $g=1$ scattering equations, even for $n=4$, and compare the resulting expressions with more standard forms of the 1-loop supergravity amplitude in terms of box integrals~\cite{Green:1982sw,Bern:1998sv}. It may be possible to make this connection already at the worldsheet level, perhaps using techniques introduced in~\cite{DHoker:1994yr} and extended in~\cite{Green:1999pv,Green:2008uj}. As a first step, it would be important to understand how to see the expected quadratic\footnote{In dimensional regularization, power law divergences do not occur, so $d=10$ type II supergravity will be accidentally finite until two loops.} UV divergence of ten dimensional supergravity. 

\medskip

The ambitwistor string formalism presented here and in~\cite{Mason:2013sva} is a chiral analogue of the usual RNS string. Thus, it comes with all the familiar shortcomings of RNS strings, such as the rather awkward spin field vertex operators, picture changing formalism and the need to sum over spin structures. In~\cite{Berkovits:2013xba}, Berkovits has presented a pure spinor version of the (gauge--fixed) ambitwistor string that, having manifest space--time supersymmetry, should in principle provide a simpler framework to study both $g=1$ and space--time supersymmetric amplitudes. It will be interesting to investigate how to compute $n$-point worldsheet correlators in these pure spinor models in closed form, perhaps using the methods of~\cite{Mafra:2011nv}.

\medskip

One of the main attractions of the amplitude representations of~\cite{Cachazo:2013hca} is that they provide such a sharp statement of the general KLT slogan of ``gravity = gauge $\times$ gauge", or better~\cite{Hodges:2011wm} ``gravity $\times$ scalar = gauge $\times$ gauge". Indeed, in~\cite{Mason:2013sva} it was also shown that a heterotic ambitwistor string theory exists whose $g=0$ correlation functions for Yang--Mills states reproduce the Yang--Mills formul\ae\ of~\cite{Cachazo:2013hca} at leading trace. It is then natural to wonder whether this relation could also be extended to higher genus. Of course, heterotic ambitwistor string amplitudes can be computed at $g=1$, and we may anticipate that they will again be localized to solutions of the $g=1$ scattering equations, and will again take the form of a Pfaffian of the matrix $M$ appearing in the supergravity calculation, times a $g=1$ current correlator. However, since the heterotic ambitwistor string also contains vertex operators corresponding to gravitational states in space--time, we do not expect these $g=1$ heterotic amplitudes to describe pure (super) Yang--Mills, even at leading trace. Thus, like the KLT relations themselves~\cite{Kawai:1985xq}, at least naively it seems that the remarkable relationships between gravity, Yang--Mills and scalars found by Cachazo {\it et al.} are restricted to tree level. In fact, the gravitational sector of the heterotic  ambitwistor string is currently rather poorly understood --- even at $g=0$, worldsheet correlators of $n$ gravitational vertex operators do not seem to agree with known formul\ae\ for gravitational scattering amplitudes. One ray of hope perhaps comes from the recent very interesting paper~\cite{Ochirov:2013xba} investigating the BCJ relations~\cite{Bern:2010ue} at higher genus in closed strings.


\medskip

\bigskip

\noindent  {\bf Acknowledgements:} The work of TA is supported by a Title A Research Fellowship at St. John's College, Cambridge. The work of EC is supported in part by the Cambridge Commonwealth, European and International Trust. The research leading to these results has received funding from the European Research Council under the European Community's Seventh Framework Programme (FP7/2007-2013) / ERC grant agreement no. [247252].

\bibliography{ATSbib}
\bibliographystyle{JHEP}

\end{document}